\documentclass{llncs}

\usepackage [utf8] {inputenc}
\usepackage {amsmath, amssymb, bussproofs, tikz, url}
\allowdisplaybreaks [4]
\usepackage{enumerate}
\usepackage{hyperref}

\title{Biometric Systems Private by Design: Reasoning about privacy properties of biometric system architectures\thanks{This work has been partially funded by the French ANR-12-INSE-0013 project BIOPRIV and the European FP7-ICT-2013-1.5 project PRIPARE. Earlier and partial versions of this work appeared in FM 2015 \cite{DBLP:conf/fm/BringerCML15} and ISC 2015 \cite{DBLP:conf/isw/BringerCML15} conferences. This work provides a global and consistent view of these preliminary publications.} 
}
\author {Julien Bringer\inst{1} \and Herv{\'e} Chabanne\inst{12} \and Daniel Le M{\'e}tayer\inst{3} \and Roch Lescuyer\inst{1}}
\institute {Safran Identity $\&$ Security, Issy-Les-Moulineaux, France
\and T{\'e}l{\'e}com ParisTech, Paris, France
\and Inria, Lyon, France}
\date {}

\begin{document}
\maketitle

\begin{abstract}
This work aims to show the applicability, and how, of privacy by design approach to biometric systems and the benefit of using formal methods to this end. Starting from a general framework that has been introduced at STM in 2014, that enables to define privacy architectures and to formally reason about their properties, we explain how it can be adapted to biometrics. 
The choice of particular techniques and the role of the components (central server, secure module, biometric terminal, smart card, \textit{etc.}) in the architecture have a strong impact on the privacy guarantees provided by a biometric system.
In the literature, some architectures have already been analysed in some way. However, the existing proposals were made on a case by case basis, which makes it difficult to compare them and to provide a rationale for the choice of specific options.
In this paper, we describe, on different architectures with various levels of protection, how a general framework for the definition of privacy architectures can be used to specify the design options of a biometric systems and to reason about them in a formal way.
\end{abstract}

\section {Introduction}
\label{s:introduction}

Applications of biometric recognition, as the most natural tool to identify or to authenticate a person, have grew over the years. They now vary from criminal investigations and identity documents to many public or private usages, like physical access control or authentication from a smartphone toward an internet service provider. Such biometric systems involve two main phases: enrolment and verification (either authentication or identification)~\cite{DBLP:journals/tcsv/JainRP04}. Enrolment is the registration phase, in which the biometric traits of a person are collected and recorded within the system.
In the {authentication} mode, a fresh biometric trait is collected and compared with the registered one by the system to check that it corresponds to the claimed identity.
In the {identification} mode, a fresh biometric data is collected and the corresponding identity is searched in a database of enrolled biometric references. 
During each phase, to enable efficient and accurate comparison, the collected biometric data are converted into discriminative features, leading to what is called a biometric template. 

With the increased use of biometric systems, and more recently with the development of personal data protection regulations, the issues related to the protection of the privacy of the used biometric traits have received particular attention. As leakage of biometric traits may lead to privacy risks, including tracking and identity theft, privacy by design approach is often needed.

As a security technical challenge, it has attracted a lot of research works since at least 15 years and a wide-array of well-documented primitives, such as encryption, homomorphic encryption, secure multi-party computation, hardware security, template protection \textit{etc.},  are known in the litterature. With those building tools, various architectures have been proposed to take into account privacy requirements in the implementation of privacy preserving biometric systems. Some solutions involve dedicated cryptographic primitives such as secure sketches~\cite{DBLP:conf/eurocrypt/DodisRS04} and fuzzy vaults~\cite{DBLP:journals/dcc/JuelsS06,DBLP:conf/avbpa/UludagPJ05}, others rely on adaptations of existing cryptographic tools~\cite{DBLP:conf/IEEEias/LiP09a} or the use of secure hardware solutions~\cite{ISOIEC24787}.
The choice of particular techniques and the role of the components (central server, secure module, terminal, smart card, \textit{etc.}) in the architecture have a strong impact on the privacy guarantees provided by a solution.
However, existing proposals were made on a case by case basis, which makes it difficult to compare them, to provide a rationale for the choice of specific options and to capitalize on past experience.

Here, we aim to show how to use and adapt a general framework that has been introduced in \cite{DBLP:conf/stm/AntignacM14} for the formal definition and validation of privacy architectures.  The goal is specify the various design options in a consistent and comparable way, and then to reason about them in a formal way in order to justify their design in terms of trust assumptions and achieved privacy properties.

The privacy by design approach is often praised by lawyers as well as computer scientists as an essential step towards a better privacy protection. It is even becoming more and more often legally compelled, as for instance in European Union with the General Data Protection Regulation~\cite{european-parliament:2014} entering into force. Nevertheless, it is one thing to impose by law the adoption of privacy by design, quite another to define precisely what it is intended to mean technically-wise and to ensure that it is put into practice by developers. The overall philosophy is that privacy should not be treated as an afterthought but rather as a first-class requirement in the design phase of systems: in other words, designers should have privacy in mind from the start when they define the features and architecture of a system. However, the practical application raises a number of challenges: first of all the privacy requirements must be defined precisely; then it must be possible to reason about potential tensions between privacy and other requirements and to explore different combinations of privacy enhancing technologies to build systems meeting all these requirements.

This work, which has been conducted in particular within the French ANR research project BioPriv \cite{Biopriv}, an interdisciplinary project involving lawyers and computer scientists, can be seen as an illustration of the feasibility of the \textit{privacy by design} approach in an industrial environment. A step in this direction has been described in \cite{DBLP:conf/stm/AntignacM14} which introduces a system for defining privacy architectures and reasoning about their properties. In Section~\ref{s:framework}, we provide an outline of this framework. Then we show how this framework can be used to apply a privacy by design approach to the implementation of biometric systems. In Sections~\ref{s:common} to~\ref{s:moc}, we describe several architectures for biometric systems, considering both existing systems and more advanced solutions, and show that they can be defined in this framework.
This makes it possible to highlight their commonalities and differences especially with regard to their underlying trust assumptions.

In the second part of this paper, we address a security issue which cannot be expressed in the framework presented in Section~\ref{s:framework}. The origin of the problem is that side-channel information may leak from the execution of the system. This issue is acute for biometric systems because the result of a matching between two biometric data inherently provides some information, even if the underlying cryptographic components are correctly implemented \cite{DBLP:conf/iih-msp/BringerCS10,DBLP:journals/tifs/SimoensBCS12,DBLP:conf/indocrypt/PagninDAM14}.
To adress this issue, in Section~\ref{s:extension}, we propose an extension of the formal framework, in which information leaks spanning over several sessions of the system can be expressed. In Section~\ref{s:mocindent}, we apply the extended model to analyse biometric information leakage in several variants of biometric system architectures.

Finally, Section~\ref{s:rel} sketches related works and Section~\ref{s:conclusion} concludes the paper with suggestions of avenues for further work.

\section {General approach}
\label{s:framework}

The work presented in \cite{DBLP:conf/stm/AntignacM14} can be seen as a first step towards a formal and systematic approach to privacy by design.
In practice, this framework makes it possible to express privacy and integrity requirements (typically the fact that an entity must obtain guarantees about the correctness of a value), to analyse their potential tensions and to make reasoned architectural choices based on explicit trust assumptions.
The motivations for the approach come from the following observations: 

\begin {itemize}
\item First, one of the key decisions that has to be taken in the design of a privacy compliant system is the location of the data and the computations: for example, a system in which all data is collected and all results computed on a central server brings strong integrity guarantees to the operator at the price of a loss of privacy for data subjects.
Decentralized solutions may provide better privacy protections but weaker guarantees for the operator.
The use of privacy enhancing technologies such as homomorphic encryption or secure multi-party computation can in some cases reconcile both objectives.

\item The choice among the architectural options should be guided by the assumptions that can be placed by the actors on the other actors and on the components of the architecture.
This trust itself can be justified in different ways (security protocol, secure or certified hardware, accredited third party, \textit{etc.}).
\end {itemize}
As far as the formal model is concerned, the framework proposed in~\cite{DBLP:conf/stm/AntignacM14} relies on a dedicated epistemic logic.
Indeed, because privacy is closely connected with the notion of knowledge, epistemic logics~\cite{fagin:2004} form an ideal basis to reason about privacy properties but standard epistemic logics based on possible worlds semantics suffer from a weakness (called ``logical omniscience''~\cite{DBLP:conf/tark/HalpernP07}) which makes them unsuitable in the context of privacy by design.

We assume that the {functionality} of the system is expressed as the computation of a set of equations $\Omega := \{{X} = T \}$ over a language $Term$ of terms~$T$ defined as follows, where $c$ represents constants ($c \in Const$), $X$ variables ($X \in Var$) and $F$ functions ($F \in Fun$):
$$
\begin {array}{ccccccccccccccccccccc}
T & ::= & {X} & | & c & | & F (T_1, \dots, T_n) 
\end {array}
$$
An {architecture} is defined by a set of components $C_i$, for $i\in[1, N]$, and a set $A$ of relations.
The relations define the capacities of the components and the trust assumptions.
We use the following language to define the relations:
$$
\begin {array} {l}
\begin {array} {rclclclcl}
A & ::= & \{R\} \\
R & ::= & Has_i (X) & | & Receive_{i,j} (\{St\}, \{X\}) & | & Compute_G (X = T) \\
 & & & | & Verify_i (St) & | & Trust_{i, j} \\
\end {array} \\ \\
\begin {array} {rclclclcl}
\makebox[6mm]{\hfill$St$} & ::= Pro & | & Att & \ Att & ::= Attest_G (\{Eq\}) \\
Pro & \multicolumn{3}{l}{::= Proof_i (\{P\})} & \ Eq & ::= Pred (T_1, \dots, T_m) \\
P & ::= Att & | & Eq \\
\end {array}
\end {array}
$$
The notation $\{Z\}$ denotes a set of terms of category $Z$.
$Has_i (X)$ denotes the fact that component $C_i$ possesses (or is the origin of) the value of $X$, which may correspond to situations in which $X$ is stored on $C_i$ or $C_i$ is a sensor collecting the value of $X$.
In this paper we use the set of predicates $Pred := \{ =, \in \}$.
$Compute_G ({X} = T)$ means that the set of components $G$ can compute the term $T$ and assign its value to $X$ and $Trust_{i, j}$ represents the fact that component $C_i$ trusts component $C_j$.
$Receive_{i,j} (\{St\}, \{X\})$ means that $C_i$ can receive the values of variables in $\{X\}$ together with the statements in $\{St\}$ from $C_j$ .

We consider two types of statements here, namely attestations: $Attest_i (\{Eq\})$ is the declaration by the component $i$ that the properties in $\{Eq\}$ hold; and proofs: $Proof_i (\{P\})$ is the delivery by $C_i$ of a set of proofs of properties.
$Verify_i$ is the verification by component $C_i$ of the corresponding statements (proof or authenticity).
In any case, the architecture level does not provide details on how a verification is done.
The verification of an attestation concerns the authenticity of the statement only, not its truth that $C_i$ may even not be able to carry out itself.
In practice, it could be the verification of a digital signature.

Graphical data flow representations can be derived from architectures expressed in this language.
For the sake of readability, we use both notations in the next sections.

The subset of the privacy logic used in this paper is the following dedicated epistemic logic:
$$
\begin {array}{cclclclclcl}
\varphi & ::= & Has_i (X) & | & Has_i^{none} (X) & | & K_i (Prop) & | & \varphi_1\ \wedge\ \varphi_2 & & \\
Prop & ::= & \multicolumn{7}{l}{Pred (T_1, \dots, T_n) \ | \ Prop_1 \ \wedge \ Prop_2} \\
\end {array}
$$
$Has_i(X)$ and $Has_i^{none}(X)$ denote the facts that component $C_i$ respectively can or cannot get the value of $X$.
$K_i$ denotes the epistemic knowledge following the ``deductive algorithmic knowledge'' philosophy~\cite{fagin:2004,DBLP:journals/logcom/Pucella06} that makes it possible to avoid the logical omniscience problem.
In this approach, the knowledge of a component $C_i$ is defined as the set of properties that this component can actually derive using its own information and his deductive system $\triangleright_i$.

Another relation, $\textit{Dep}_i$, is used to take into account dependencies between variables.
$\textit{Dep}_i (Y, \mathcal{X})$ means that if $C_i$ can obtain the values of each variable in the set of variables $\mathcal{X}$, then it may be able to derive the value of $Y$.
The absence of such a relation is an assumption that $C_i$ cannot derive the value of $X$ from the values of the variables in $\mathcal{X}$.
It should be noted that this dependency relation is associated with a given component: different components may have different capacities.
For example, if component $C_i$ is the only component able to decrypt a variable $ev$ to get the clear text $v$, then $\textit{Dep}_i (v, \{ ev \} )$ holds but $\textit{Dep}_j (v,\{ ev\} )$ does not hold for any $j \neq i$.

The semantics ${S} (A)$ of an architecture $A$ is defined as the set of states of the components $C_i$ of $A$ resulting from compatible execution traces~\cite{DBLP:conf/stm/AntignacM14}.
A compatible execution trace contains only events that are instantiations of relations (e.g. $Receive_{i,j}, Compute_G$, \textit{etc.}) of $A$ (as further discussed in Section \ref{s:extarch}).
The semantics $S (\varphi)$ of a property $\varphi$ is defined as the set of architectures meeting $\varphi$.
For example, $A \in S (Has_i^{none} (X))$ if for all states $\sigma \in S (A)$, the sub-state $\sigma_i$ of component $C_i$ is such that $\sigma_i (X) = \bot$, which expresses the fact that the component $C_i$ cannot assign a value to the variable $X$.

To make it possible to reason about privacy properties, an axiomatics of this logic is presented and is proven sound and complete.
$A \vdash \varphi$ denotes that $\varphi$ can be derived from $A$ thanks to the deductive rules (\textit{i.e.} there exists a derivation tree such that all steps belong to the axiomatics, and such that the leaf is $A \vdash \varphi$).
A subset of the axioms useful for this paper is presented in Figure~\ref{fig:axiomatics}.

\begin {figure}[t!]
\begin {center}
\fbox {\begin {minipage}{.98\textwidth}

\AxiomC {$Has_i (X) \in A$}
\LeftLabel {\textsf{H1}}
\UnaryInfC {$A \vdash Has_i (X)$}
\DisplayProof
\qquad
\AxiomC {$Compute_G (X = T) \in A$}
\AxiomC {$C_i \in G$}
\LeftLabel {\textsf{H3}}
\BinaryInfC {$A \vdash Has_i (X)$}
\DisplayProof

\smallskip

\AxiomC {$Receive_{i,j} (S, E) \in A$}
\AxiomC {$X \in E$}
\LeftLabel {\textsf{H2}}
\BinaryInfC {$A \vdash Has_i (X)$}
\DisplayProof
\qquad
\AxiomC {$Dep_i (Y, \mathcal{X})$}
\AxiomC {$\forall X \in \mathcal{X}, A \vdash Has_i (X)$}
\LeftLabel {\textsf{H5}}
\BinaryInfC {$A \vdash Has_i (Y)$}
\DisplayProof

\smallskip

\smallskip

\AxiomC {$A \nvdash Has_i (X)$}
\LeftLabel{\textsf{HN}}
\UnaryInfC {$A \vdash Has^{none}_i (X)$}
\DisplayProof
\qquad
\AxiomC {$E \triangleright_i Eq_0$}
\AxiomC {$\forall Eq \in E: A \vdash K_i (Eq)$}
\LeftLabel {\textsf{K}$\mathbf{\triangleright}$}
\BinaryInfC {$A \vdash K_i (Eq_0)$}
\DisplayProof

\smallskip

\AxiomC {$Compute_G (X = T) \in A$
\quad\,
$C_i \in G$}
\LeftLabel {\textsf{K1}}
\UnaryInfC {$A \vdash K_i (X = T)$}
\DisplayProof
\qquad
\AxiomC {$Verify_i (Proof_j (E)) \in A$}
\AxiomC {$Eq \in E$}
\LeftLabel {\textsf{K3}}
\BinaryInfC {$A \vdash K_i (Eq)$}
\DisplayProof

\smallskip

\AxiomC {$Verify_i (Proof_j (E)) \in A$}
\AxiomC {$Attest_k (E') \in E$}
\AxiomC {$Eq \in E'$}
\AxiomC {$Trust_{i, k} \in A$}
\LeftLabel {\textsf{K4}}
\QuaternaryInfC {$A \vdash K_i (Eq)$}
\DisplayProof

\smallskip

\AxiomC {$Verify_i (Attest_j (E)) \in A$}
\AxiomC {$Trust_{i, j} \in A$}
\AxiomC {$Eq \in E$}
\LeftLabel {\textsf{K5}}
\TrinaryInfC {$A \vdash K_i (Eq)$}
\DisplayProof
\end {minipage}}
\end {center}
\caption{A subset of rules from the axiomatics of \cite{DBLP:conf/stm/AntignacM14}}
\label{fig:axiomatics}
\end {figure}

\section {Biometric systems architectures}
\label{s:common}

Before starting the presentation of the different biometric architectures in the next sections, we introduce in this section the basic terminology used in this paper and the common features of the architectures.
For the sake of readability, we use upper case sans serif letters \textsf{S}, \textsf{T}, \textit{etc.} rather than indexed variables $C_i$ to denote components.
By abuse of notation, we will use component names instead of indices and write, for example, $Receive_{\mathsf{U}, \mathsf{T}} (\{\}, \{\mathtt{dec}\})$.
Type letters \texttt{dec}, \texttt{br}, \textit{etc.} denote variables.
The set of components of an architecture is denoted by $\mathcal{J}$.

\begin {figure} [ht]
\begin {center}
\fbox {\begin {minipage}{.98\textwidth}
\begin {center}
\begin {tikzpicture} [scale=0.06]

\node at (-80,0)
{
\begin {tikzpicture} [scale=0.1]
\draw (0,10) circle (2);
\draw (0,8) -- (0,4);
\draw (0,4) -- (-2,0);
\draw (0,4) -- (2,0);
\draw (0,6) -- (-2,6);
\draw (0,6) -- (2,6);
\end {tikzpicture}
};
\node at (-80,-16) {User};
\node [scale=0.6] at (-40,0)
{
\begin {tikzpicture} [scale=0.1]
\draw (0,-10) ellipse (7cm and 3cm);
\draw [white,fill=white] (0,-9.5) ellipse (7cm and 3cm);
\draw (0,10) ellipse (7cm and 3cm);
\draw (7,10) -- (7,-10);
\draw (-7,10) -- (-7,-10);
\draw [fill=yellow, even odd rule, yshift=-9cm, xshift=1cm] (0,0) rectangle (4,4) (1.6,2.4) -- (1.4,0.6) -- (2.6,0.6) -- (2.4,2.4) arc (-40:220:0.54);
\draw [fill=yellow, yshift=-9cm, xshift=1cm] (4,4) -- (4,5) arc (0:180:2) -- (0,4) -- (0.6,4) -- (0.6,5) arc (180:0:1.4) -- (3.4,4) -- (4,4);
\end {tikzpicture}
};
\node at (-40,-16) {Encrypted};
\node at (-40,-21) {database};
\node [scale=0.3] at (0,0)
{
\begin {tikzpicture} [scale=0.5]
\draw [rounded corners] (0,0) rectangle (8,10);
\node at (6.5,4) {\includegraphics [width=1.1cm] {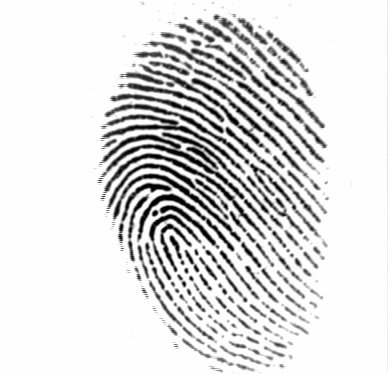}};
\draw (5.5,2.5) rectangle (7.5,5.5);
\draw [rotate=-30,fill=white] (1,0) -- (1,4) arc (180:0:1 and 1.3) -- (3,0);
\draw [rotate=-30,fill=white] (2,4) ellipse (0.8 and 1.4);
\end {tikzpicture}
};

\node at (0,-16) {Terminal};
\node [scale=0.3] at (40,0)
{
\begin {tikzpicture} [scale=0.5]
\draw [rounded corners] (0,0) rectangle (10,6);
\draw [fill=yellow, rounded corners] (1,3) rectangle (2.2,4.2);
\draw (1.4,3) -- (1.4,4.2);
\draw (1.8,3) -- (1.8,4.2);
\draw (1,3.3) -- (2.2,3.3);
\draw (1,3.6) -- (2.2,3.6);
\draw (1,3.9) -- (2.2,3.9);
\end {tikzpicture}
};

\node at (40,-16) {Card};
\draw [rounded corners, red, dashed,thick] (75,-5) rectangle (85,5);
\node at (80,-11) {Location};
\node at (80,-16) {of the};
\node at (80,-21) {comparison};

\end {tikzpicture}
\end {center}
\end {minipage}}
\end {center}
\caption {Graphical representations}
\label{fig:legend}
\end {figure}

The variables used in biometric system architectures are the following:

\begin {itemize}

\item A biometric reference template \texttt{br} built during the enrolment phase, where a template corresponds to a set or vector of biometrics features that are extracted from raw biometric data in order to be able to compare biometric data accurately.
\item A raw biometric data \texttt{rd} provided by the user during the verification phase.
\item A fresh template \texttt{bs} derived from \texttt{rd} during the verification phase.
\item A threshold \texttt{thr} which is used during the verification phase as a closeness criterion for the biometric templates.
\item The output \texttt{dec} of the verification which is the result of the matching between the fresh template \texttt{bs} and the enrolled templates \texttt{br}, considering the threshold \texttt{thr}.

\end {itemize} 
Two components appear in all biometric architectures: a component $\mathsf{U}$ representing the user, and the terminal \textsf{T} which is equipped with a sensor used to acquire biometric traits.
In addition, biometric architectures may involve an explicit issuer~\textsf{I}, enrolling users and certifying their templates, a server \textsf{S} managing a database containing enrolled templates, a module (which can be a hardware security module, denoted HSM) to perform the matching and eventually to take the decision, and a smart card \textsf{C} to store the enrolled templates (and in some cases to perform the matching).
Figure~\ref{fig:legend} introduces some graphical representations used in the figures of this paper.

In this paper, we focus on the verification phase and assume that enrolment has already been done.
Therefore the biometric reference templates are stored on a component which can be either the issuer ($Has_\mathsf{I} (\mathtt{br})$) or a smart card ($Has_\mathsf{C} (\mathtt{br})$).
A verification process is initiated by the terminal \textsf{T} receiving as input a raw biometric data \texttt{rd} from the user \textsf{U}.
\textsf{T} extracts the fresh biometric template \texttt{bs} from \texttt{rd} using the function $Extract \in Fun$.
All architectures $A$ therefore include $ Receive_{\mathsf{T}, \mathsf{U}} (\{\}, \{\mathtt{rd}\}) $ and $ Compute_\mathsf{T} (\mathtt{bs} = Extract (\mathtt{rd}))$ and the $Dep_\mathsf{T}$ relation is such that $(\mathtt{bs}, \{\mathtt{rd}\}) \in Dep_\mathsf{T}$.
In all architectures $A$, the user receives the final decision $\mathtt{dec}$ (which can typically be positive or negative) from the terminal: $Receive_{\mathsf{U}, \mathsf{T}} (\{\}, \{\mathtt{dec}\}) \in A$.
The matching itself, which can be performed by different components depending on the architecture, is expressed by the function $\mu \in Fun$ which takes as arguments two biometric templates and the threshold~$\mathtt{thr}$.

\section{Application of the framework to  several architectures for biometric systems with various protection levels}
\subsection {Protecting the reference templates with encryption}
\label{s:proctemp}

Let us consider first the most common architecture deployed for protecting biometric data.
When a user is enrolled his reference template is stored encrypted, either in a terminal with an embedded database, or in a central database.
During the identification process, the user supplies a fresh template, the reference templates are decrypted by a component (which can be typically the terminal or a dedicated hardware security module) and the comparison is done inside this component.
The first part of Figure~\ref{fig:hsm} shows an architecture $A_\mathsf{ed}$ in which reference templates are stored in a central database and the decryption of the references and the matching are done inside the terminal.
The second part of the figure shows an architecture $A_\mathsf{hsm} $ in which the decryption of the references and the matching are done on a dedicated hardware security module.
Both architectures are considered in turn in the following paragraphs.

\begin {figure} [ht]
\begin {center}
\fbox {\begin {minipage}{.98\textwidth} \small
\begin {center}
\begin {tikzpicture}[scale=0.09]

\node [above] at (-54,-17) {$\mathsf{U}$};
\node [scale=1.2] at (-54,0)
{
\begin {tikzpicture} [scale=0.1]
\draw (0,10) circle (2);
\draw (0,8) -- (0,4);
\draw (0,4) -- (-2,0);
\draw (0,4) -- (2,0);
\draw (0,6) -- (-2,6);
\draw (0,6) -- (2,6);
\end {tikzpicture}
};

\draw [->] (-50,2) -- (-30,2);
\node [above] at (-40,2) {$\mathtt{rd}$};
\draw [<-] (-50,-2) -- (-30,-2);
\node [below] at (-40,-2) {$\mathtt{dec}$};

\node [above] at (-18,-17) {$\mathsf{T}$};
\node [scale=0.4] at (-18,0)
{
\begin {tikzpicture} [scale=0.5]
\draw [rounded corners] (0,0) rectangle (8,10);
\node at (6.5,4) {\includegraphics [width=1.1cm] {inputs/1101_1.jpg}};
\draw (5.5,2.5) rectangle (7.5,5.5);
\draw [rotate=-30,fill=white] (1,0) -- (1,4) arc (180:0:1 and 1.3) -- (3,0);
\draw [rotate=-30,fill=white] (2,4) ellipse (0.8 and 1.4);
\end {tikzpicture}
};

\node at (-18,10) {$\mathtt{rd} \to \mathtt{bs}$};
\node at (-18,6) {$\mathtt{thr}$};
\draw [rounded corners, red, dashed,thick] (-28,-10.5) rectangle (-8,14);

\draw [<-] (-6,0) -- (10,0);
\node [above] at (2,0) {$\mathtt{ebr}$};

\node [above] at (18,-17) {$\mathsf{S}$};
\node [scale=0.8] at (18,0)
{
\begin {tikzpicture} [scale=0.1]
\draw (0,-10) ellipse (7cm and 3cm);
\draw [white,fill=white] (0,-9.5) ellipse (7cm and 3cm);
\draw (0,10) ellipse (7cm and 3cm);
\draw (7,10) -- (7,-10);
\draw (-7,10) -- (-7,-10);
\draw [fill=yellow, even odd rule, yshift=-9cm, xshift=1cm] (0,0) rectangle (4,4) (1.6,2.4) -- (1.4,0.6) -- (2.6,0.6) -- (2.4,2.4) arc (-40:220:0.54);
\draw [fill=yellow, yshift=-9cm, xshift=1cm] (4,4) -- (4,5) arc (0:180:2) -- (0,4) -- (0.6,4) -- (0.6,5) arc (180:0:1.4) -- (3.4,4) -- (4,4);
\end {tikzpicture}
};


\draw [<-] (26,0) -- (43,0);
\node [above] at (34.5,0) {$\mathtt{ebr}$};

\node [above] at (54,-17) {$\mathsf{I}$};
\node [scale=0.4] at (54,0)
{
\begin {tikzpicture} [scale=0.5]
\draw [rounded corners] (0,0) rectangle (8,10);
\node at (1.5,6) {\includegraphics [width=1.1cm] {inputs/1101_1.jpg}};
\draw (0.5,4.5) rectangle (2.5,7.5);
\end {tikzpicture}
};

\node at (54,8) {$\mathtt{br} \to \mathtt{ebr}$};

\end {tikzpicture}
\end {center}
\end {minipage}}

\smallskip

Encrypted database

\smallskip

\fbox {\begin {minipage}{.98\textwidth} \small
\begin {center}
\begin {tikzpicture}[scale=0.09]

\node [above] at (-56,-17) {$\mathsf{U}$};
\node [scale=1.2] at (-56,0)
{
\begin {tikzpicture} [scale=0.1]
\draw (0,10) circle (2);
\draw (0,8) -- (0,4);
\draw (0,4) -- (-2,0);
\draw (0,4) -- (2,0);
\draw (0,6) -- (-2,6);
\draw (0,6) -- (2,6);
\end {tikzpicture}
};

\draw [->] (-52,2) -- (-40,2);
\node [above] at (-46,2) {$\mathtt{rd}$};
\draw [<-] (-52,-2) -- (-40,-2);
\node [below] at (-46,-2) {$\mathtt{dec}$};

\node [above] at (-28,-17) {$\mathsf{T}$};
\node [scale=0.4] at (-28,0)
{
\begin {tikzpicture} [scale=0.5]
\draw [rounded corners] (0,0) rectangle (8,10);
\node at (6.5,4) {\includegraphics [width=1.1cm] {inputs/1101_1.jpg}};
\draw (5.5,2.5) rectangle (7.5,5.5);
\draw [rotate=-30,fill=white] (1,0) -- (1,4) arc (180:0:1 and 1.3) -- (3,0);
\draw [rotate=-30,fill=white] (2,4) ellipse (0.8 and 1.4);
\end {tikzpicture}
};

\node at (-28,10) {$\mathtt{rd} \to \mathtt{bs}$};

\draw [->] (-18,-3) -- (-6,-3);
\node [above] at (-12,-3) {$\mathtt{bs}, \mathtt{ebr}$};
\draw [<-] (-18,-7) -- (-6,-7);
\node [below] at (-12,-7) {$\mathtt{dec}$};

\node [above] at (7,-17) {$\mathsf{M}$};
\draw [rounded corners] (-3,-10) rectangle (17,0);
\node at (7,-5) {$\mathtt{thr}$};
\draw [rounded corners, red, dashed,thick] (-4,-11) rectangle (18,1);

\draw [<-] (-18,6) -- (20,6);
\node [above] at (1,6) {$\mathtt{ebr}$};

\node [above] at (28,-17) {$\mathsf{S}$};
\node [scale=0.8] at (28,0)
{
\begin {tikzpicture} [scale=0.1]
\draw (0,-10) ellipse (7cm and 3cm);
\draw [white,fill=white] (0,-9.5) ellipse (7cm and 3cm);
\draw (0,10) ellipse (7cm and 3cm);
\draw (7,10) -- (7,-10);
\draw (-7,10) -- (-7,-10);
\draw [fill=yellow, even odd rule, yshift=-9cm, xshift=1cm] (0,0) rectangle (4,4) (1.6,2.4) -- (1.4,0.6) -- (2.6,0.6) -- (2.4,2.4) arc (-40:220:0.54);
\draw [fill=yellow, yshift=-9cm, xshift=1cm] (4,4) -- (4,5) arc (0:180:2) -- (0,4) -- (0.6,4) -- (0.6,5) arc (180:0:1.4) -- (3.4,4) -- (4,4);
\end {tikzpicture}
};


\draw [<-] (36,0) -- (45,0);
\node [above] at (40.5,0) {$\mathtt{ebr}$};

\node [above] at (56,-17) {$\mathsf{I}$};
\node [scale=0.4] at (56,0)
{
\begin {tikzpicture} [scale=0.5]
\draw [rounded corners] (0,0) rectangle (8,10);
\node at (1.5,6) {\includegraphics [width=1.1cm] {inputs/1101_1.jpg}};
\draw (0.5,4.5) rectangle (2.5,7.5);
\end {tikzpicture}
};

\node at (56,8) {$\mathtt{br} \to \mathtt{ebr}$};

\end {tikzpicture}
\end {center}
\end {minipage}}

\smallskip

Encrypted database with a hardware security module (HSM)
\end {center}
\caption {Classical architectures with an encrypted database}
\label{fig:hsm}
\end {figure}

\subsubsection{Use of an encrypted database.}
The first architecture $A_\mathsf{ed}$ is composed of a user \textsf{U}, a terminal \textsf{T}, a server \textsf{S} managing an encrypted database \texttt{ebr} and
an issuer \textsf{I} enrolling users and generating the encrypted database \texttt{ebr}.
The set $Fun$ includes the encryption and decryption functions $Enc$ and $Dec$.
When applied to an array, $Enc$ is assumed to encrypt each entry of the array.
At this stage, for the sake of conciseness, we consider only biometric data in the context of an identification phase.
The same types of architectures can be used to deal with authentication, which does not raise any specific issue.
The functionality of the architecture is
$\Omega :=$ \{$\mathtt{ebr} = Enc (\mathtt{br})$,
$\mathtt{br}' = Dec (\mathtt{ebr})$,
$\mathtt{bs} = Extract (\mathtt{rd})$,
$\mathtt{dec} = \mu (\mathtt{br}', \mathtt{bs}, \mathtt{thr})$\},
and the architecture is defined as:
\begin {align*}
A_\mathsf{ed} := \big\{
&
Has_\mathsf{I} (\mathtt{br}), Has_\mathsf{U} (\mathtt{rd}), Has_\mathsf{T} (\mathtt{thr}),
Compute_\mathsf{I} (\mathtt{ebr} = Enc (\mathtt{br})),
\\
&
Receive_{\mathsf{S}, \mathsf{I}} (\{Attest_\mathsf{I} (\mathtt{ebr} = Enc (\mathtt{br}))\}, \{\mathtt{ebr}\}),
\\
&
Receive_{\mathsf{T}, \mathsf{S}} (\{Attest_\mathsf{I} (\mathtt{ebr} = Enc (\mathtt{br}))\}, \{\mathtt{ebr}\}),
Trust_{\mathsf{T}, \mathsf{I}},
\\
&
Verify_\mathsf{T} (Attest_\mathsf{I} (\mathtt{ebr} = Enc (\mathtt{br}))),
Receive_{\mathsf{T}, \mathsf{U}} (\{\}, \{\mathtt{rd}\}),
\\
&
Compute_\mathsf{T} (\mathtt{bs} = Extract (\mathtt{rd})),
Compute_\mathsf{T} (\mathtt{br}' = Dec (\mathtt{ebr})),
\\
&
Compute_\mathsf{T} (\mathtt{dec} = \mu (\mathtt{br}', \mathtt{bs}, \mathtt{thr})),
Receive_{\mathsf{U}, \mathsf{T}} (\{\}, \{\mathtt{dec}\})
\big\}
\end {align*}
The properties of the encryption scheme are captured by the dependence and deductive relations.
The dependence relations are:
$(\mathtt{ebr}, \{\mathtt{br}\}) \in Dep_\mathsf{I}$, and
\{($\mathtt{bs}$, \{$\mathtt{rd}$\}), ($\mathtt{dec}$,
\{$\mathtt{br'}$, $\mathtt{bs}$, $\mathtt{thr}$\}),
($\mathtt{br}'$, \{$\mathtt{ebr}$\}), ($\mathtt{br}$, \{$\mathtt{ebr}$\})\} $\subseteq$ $Dep_\mathsf{T}$.
Moreover the deductive algorithm relation contains:
$\{\mathtt{ebr} = Enc (\mathtt{br})\} \triangleright \{\mathtt{br} = Dec (\mathtt{ebr})\}$.

From the point of view of biometric data protection, the property that this architecture is meant to ensure is the fact that the server should not have access to the reference template, that is to say: $Has_\mathsf{S}^{none} (\mathtt{br})$, which can be proven using 
Rule \textsf{HN} (the same property holds for $\mathtt{br}'$):

\begin {center}
\AxiomC {$Has_{\mathsf{S}} (\mathtt{br}) \not\in A_\mathsf{ed}$
\quad
$\nexists \mathcal{X}: (\mathtt{br}, \mathcal{X}) \in Dep_\mathsf{S}$
\quad
$\nexists T: Compute_\mathsf{S} (\mathtt{br} = T) \in A_\mathsf{ed}$}
\noLine
\UnaryInfC {$\nexists j \in \mathcal{J}, \nexists S, \nexists E, Receive_{\mathsf{S}, j} (S, E) \in A_\mathsf{ed} \wedge \mathtt{br} \in E$}
\LeftLabel {\textsf{HN}}
\UnaryInfC {$A_\mathsf{ed} \vdash Has_\mathsf{S}^{none} (\mathtt{br})$}
\DisplayProof
\end {center}
It is also easy to prove, using \textsf{H2} and \textsf{H5}, that the terminal has access to $\mathtt{br}'$: $Has_\mathsf{T} (\mathtt{br}')$.

As far as integrity is concerned, the terminal should be convinced that the matching is correct.
The proof relies on the trust placed by the terminal in the issuer (about the correctness of $\mathtt{ebr}$) and the computations that the terminal can perform by itself (through $Compute_\mathsf{T}$ and the application of $\triangleright$):

\begin {center}
\AxiomC {$Verify_\mathsf{T} (\{Attest_\mathsf{I} (\mathtt{ebr} = Enc (\mathtt{br}))\}) \in A_\mathsf{ed}$}
\AxiomC {$Trust_{\mathsf{T}, \mathsf{I}} \in A_\mathsf{ed}$}
\LeftLabel {\textsf{K5}}
\BinaryInfC {$A_\mathsf{ed} \vdash K_\mathsf{T} (\mathtt{ebr} = Enc (\mathtt{br}))$}
\DisplayProof

\medskip

\AxiomC {$\{\mathtt{ebr} = Enc (\mathtt{br})\} \triangleright \{\mathtt{br} = Dec (\mathtt{ebr})\}$}
\AxiomC {$A_\mathsf{ed} \vdash K_\mathsf{T} (\mathtt{ebr} = Enc (\mathtt{br}))$}
\LeftLabel {\textsf{K$\triangleright$}}
\BinaryInfC {$A_\mathsf{ed} \vdash K_\mathsf{T} (\mathtt{br} = Dec (\mathtt{ebr}))$}
\DisplayProof

\medskip

\AxiomC {$Compute_\mathsf{T} (\mathtt{br}' = Dec (\mathtt{ebr})) \in A_\mathsf{ed}$}
\LeftLabel {\textsf{K1}}
\UnaryInfC {$A_\mathsf{ed} \vdash K_\mathsf{T} (\mathtt{br}' = Dec (\mathtt{ebr}))$}
\DisplayProof
\end {center}
Assuming that all deductive relations include the properties (commutativity and transitivity) of the equality, \textsf{K$\triangleright$} can be used to derive:
$A_\mathsf{ed} \vdash K_\mathsf{T} (\mathtt{br} = \mathtt{br}')$.
A further application of \textsf{K1} with another transitivity rule for the equality allows us to obtain the desired integrity property:

\begin {center}
\AxiomC {$A_\mathsf{ed} \vdash K_\mathsf{T} (\mathtt{br} = \mathtt{br}')$}
\AxiomC {$Compute_\mathsf{T} (\mathtt{dec} = \mu (\mathtt{br}', \mathtt{bs}, \mathtt{thr})) \in A_\mathsf{ed}$}
\LeftLabel {\textsf{K1}}
\UnaryInfC {$A_\mathsf{ed} \vdash K_\mathsf{T} (\mathtt{dec} = \mu (\mathtt{br}', \mathtt{bs}, \mathtt{thr}))$}
\LeftLabel {\textsf{K$\triangleright$}}
\BinaryInfC {$A_\mathsf{ed} \vdash K_\mathsf{T} (\mathtt{dec} = \mu (\mathtt{br}, \mathtt{bs}, \mathtt{thr}))$}
\DisplayProof
\end {center}

\subsubsection {Encrypted database with a hardware security module.}
The architecture presented in the previous subsection relies on the terminal to decrypt the reference template and to perform the matching operation.
As a result, the clear reference template is known by the terminal and the only component that has to be trusted by the terminal is the issuer.
If it does not seem sensible to entrust the terminal with this central role, another option is to delegate the decryption of the reference template and computation of the matching to a hardware security module so that the terminal itself never stores any clear reference template.
This strategy leads to architecture $A_\mathsf{hsm}$ pictured in the second part of Figure~\ref{fig:hsm}.

In addition to the user \textsf{U}, the issuer \textsf{I}, the terminal \textsf{T}, and the server \textsf{S}, the set of components contains a hardware security module \textsf{M}.
The terminal does not perform the matching, but has to trust \textsf{M}.
This trust can be justified in practice by the level of security provided by the HSM \textsf{M} (which can also be endorsed by an official security certification scheme).
The architecture is described as follows in our framework:
\begin {align*}
A_\mathsf{hsm} := \big\{
&
Has_\mathsf{I} (\mathtt{br}), Has_\mathsf{U} (\mathtt{rd}), Has_\mathsf{M} (\mathtt{thr}),
Compute_\mathsf{I} (\mathtt{ebr} = Enc (\mathtt{br})),
\\
&
Receive_{\mathsf{S}, \mathsf{I}} (\{Attest_\mathsf{I} (\mathtt{ebr} = Enc (\mathtt{br}))\}, \{\mathtt{ebr}\}),
\\
&
Receive_{\mathsf{T}, \mathsf{S}} (\{Attest_\mathsf{I} (\mathtt{ebr} = Enc (\mathtt{br}))\}, \{\mathtt{ebr}\}),
Trust_{\mathsf{T}, \mathsf{I}},
\\
&
Verify_\mathsf{T} (Attest_\mathsf{I} (\mathtt{ebr} = Enc (\mathtt{br}))),
Receive_{\mathsf{T}, \mathsf{U}} (\{\}, \{\mathtt{rd}\}),
\\
&
Compute_\mathsf{T} (\mathtt{bs} = Extract (\mathtt{rd})),
Receive_{\mathsf{M}, \mathsf{T}} (\{\}, \{\mathtt{bs}, \mathtt{ebr}\}),
\\
&
Compute_\mathsf{M} (\mathtt{br}' = Dec (\mathtt{ebr})),
Compute_\mathsf{M} (\mathtt{dec} = \mu (\mathtt{br}', \mathtt{bs}, \mathtt{thr})),
\\
&
Verify_\mathsf{T} (\{Attest_\mathsf{M} (\mathtt{dec} = \mu (\mathtt{br}', \mathtt{bs}, \mathtt{thr}))\}),
Trust_{\mathsf{T}, \mathsf{M}},
\\
&
Receive_{\mathsf{T}, \mathsf{M}} (\mathcal{A}, \{\mathtt{dec}\}),
Verify_\mathsf{T} (\{Attest_\mathsf{M} (\mathtt{br}' = Dec (\mathtt{ebr}))\})
\big\}
\end {align*}
where the set of attestations $\mathcal{A}$ received by the terminal from the module is
$\mathcal{A} := \{Attest_\mathsf{M} (\mathtt{dec} = \mu (\mathtt{br}', \mathtt{bs}, \mathtt{thr})), Attest_\mathsf{M} (\mathtt{br}' = Dec (\mathtt{ebr}))\}$.

The trust relation between the terminal and the module makes it possible to apply rule \textsf{K5} twice:

\begin {center}
\AxiomC {$Verify_\mathsf{T} (\{Attest_\mathsf{M} (\mathtt{dec} = \mu (\mathtt{br}', \mathtt{bs}, \mathtt{thr}))\}) \in A_\mathsf{hsm}$
\quad
$Trust_{\mathsf{T}, \mathsf{M}} \in A_\mathsf{hsm}$}
\UnaryInfC {$A_\mathsf{hsm} \vdash K_\mathsf{T} (\mathtt{dec} = \mu (\mathtt{br}', \mathtt{bs}, \mathtt{thr}))$}
\DisplayProof

\smallskip

\AxiomC {$Verify_\mathsf{T} (\{Attest_\mathsf{M} (\mathtt{br'} = Dec (\mathtt{ebr}))\}) \in A_\mathsf{hsm}$}
\AxiomC {$Trust_{\mathsf{T}, \mathsf{M}} \in A_\mathsf{hsm}$}
\LeftLabel {\textsf{K5}}
\BinaryInfC {$A_\mathsf{hsm} \vdash K_\mathsf{T} (\mathtt{br'} = Dec (\mathtt{ebr}))$}
\DisplayProof
\end {center}
The same proof as in the previous subsection can be applied to establish the integrity of the matching.
The trust relation between the terminal and the issuer and the rules \textsf{K5}, \textsf{K$\triangleright$} make it possible to derive:
$A_\mathsf{hsm} \vdash K_\mathsf{T} (\mathsf{br} = Dec (\mathsf{ebr}))$.
Then two successive applications of \textsf{K$\triangleright$} regarding the transitivity of the equality lead to:
$A_\mathsf{hsm} \vdash K_\mathsf{T} (\mathtt{dec} = \mu (\mathtt{br}, \mathtt{bs}, \mathtt{thr}))$.

As in architecture $A_\mathsf{ed} $, the biometric references are never disclosed to the server.
However, in contrast with $A_\mathsf{ed} $, they are not disclosed either to the terminal, as shown by rule \textsf{HN}:

\begin {center}
\AxiomC {$Has_{\mathsf{T}} (\mathtt{br}) \not\in A_\mathsf{hsm}$
\ \ 
$\nexists \mathcal{X}: (\mathtt{br}, \mathcal{X}) \in Dep_\mathsf{T}$
\ \ 
$\nexists T: Compute_\mathsf{T} (\mathtt{br} = T) \in A_\mathsf{hsm}$}
\noLine
\UnaryInfC {\makebox [.9\textwidth] {$\nexists j \in \mathcal{J}, \nexists S, \nexists E, Receive_{\mathsf{T}, j} (S, E) \in A_\mathsf{hsm} \wedge \mathtt{br} \in E$}}
\LeftLabel {\textsf{HN}}
\UnaryInfC {$A_\mathsf{hsm} \vdash Has_\mathsf{T}^{none} (\mathtt{br})$}
\DisplayProof
\end {center}

\subsection {Enhancing protection with homomorphic encryption}
\label{s:homenc}

In both architectures of Section~\ref{s:proctemp}, biometric templates are protected, but the component performing the matching (either the terminal or the secure module) gets access to the reference templates.
In this section, we show how homomorphic encryption can be used to ensure that no component gets access to the biometric reference templates during the verification.

\begin {figure} [ht]
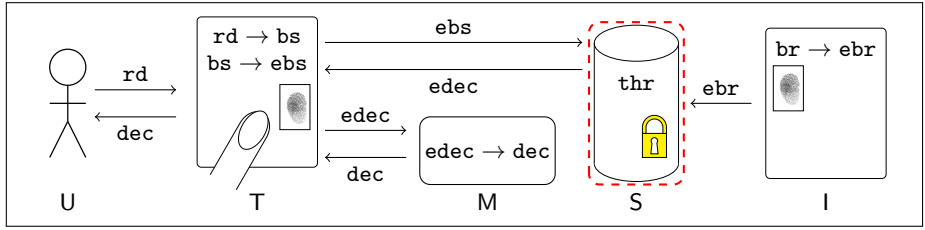

\begin {center}
\fbox {\begin {minipage}{.98\textwidth} \small
\begin {center}
\begin {tikzpicture}[scale=0.09]

\node [above] at (-56,-17) {$\mathsf{U}$};
\node [scale=1.2] at (-56,0)
{
\begin {tikzpicture} [scale=0.1]
\draw (0,10) circle (2);
\draw (0,8) -- (0,4);
\draw (0,4) -- (-2,0);
\draw (0,4) -- (2,0);
\draw (0,6) -- (-2,6);
\draw (0,6) -- (2,6);
\end {tikzpicture}
};

\draw [->] (-52,2) -- (-40,2);
\node [above] at (-46,2) {$\mathtt{rd}$};
\draw [<-] (-52,-2) -- (-40,-2);
\node [below] at (-46,-2) {$\mathtt{dec}$};

\node [above] at (-28,-17) {$\mathsf{T}$};
\node [scale=0.4] at (-28,0)
{
\begin {tikzpicture} [scale=0.5]
\draw [rounded corners] (0,0) rectangle (8,10);
\node at (6.5,4) {\includegraphics [width=1.1cm] {inputs/1101_1.jpg}};
\draw (5.5,2.5) rectangle (7.5,5.5);
\draw [rotate=-30,fill=white] (1,0) -- (1,4) arc (180:0:1 and 1.3) -- (3,0);
\draw [rotate=-30,fill=white] (2,4) ellipse (0.8 and 1.4);
\end {tikzpicture}
};

\node at (-28,10) {$\mathtt{rd} \to \mathtt{bs}$};
\node at (-28,6) {$\mathtt{bs} \to \mathtt{ebs}$};

\draw [->] (-18,-4) -- (-6,-4);
\node [above] at (-12,-4) {$\mathtt{edec}$};
\draw [<-] (-18,-8) -- (-6,-8);
\node [below] at (-12,-8) {$\mathtt{dec}$};

\node [above] at (6,-17) {$\mathsf{M}$};
\draw [rounded corners] (-4,-12) rectangle (16,-2);
\node at (6,-7) {$\mathtt{edec} \to \mathtt{dec}$};

\draw [->] (-18,9) -- (20,9);
\node [above] at (1,9) {$\mathtt{ebs}$};
\draw [<-] (-18,5) -- (20,5);
\node [below] at (1,5) {$\mathtt{edec}$};

\node [above] at (28,-17) {$\mathsf{S}$};
\node [scale=0.8] at (28,0)
{
\begin {tikzpicture} [scale=0.1]
\draw (0,-10) ellipse (7cm and 3cm);
\draw [white,fill=white] (0,-9.5) ellipse (7cm and 3cm);
\draw (0,10) ellipse (7cm and 3cm);
\draw (7,10) -- (7,-10);
\draw (-7,10) -- (-7,-10);
\draw [fill=yellow, even odd rule, yshift=-9cm, xshift=1cm] (0,0) rectangle (4,4) (1.6,2.4) -- (1.4,0.6) -- (2.6,0.6) -- (2.4,2.4) arc (-40:220:0.54);
\draw [fill=yellow, yshift=-9cm, xshift=1cm] (4,4) -- (4,5) arc (0:180:2) -- (0,4) -- (0.6,4) -- (0.6,5) arc (180:0:1.4) -- (3.4,4) -- (4,4);
\end {tikzpicture}
};

\node at (28,3) {$\mathtt{thr}$};
\draw [rounded corners, red, dashed,thick] (21,-12) rectangle (35,12);

\draw [<-] (36,0) -- (45,0);
\node [above] at (40.5,0) {$\mathtt{ebr}$};

\node [above] at (56,-17) {$\mathsf{I}$};
\node [scale=0.4] at (56,0)
{
\begin {tikzpicture} [scale=0.5]
\draw [rounded corners] (0,0) rectangle (8,10);
\node at (1.5,6) {\includegraphics [width=1.1cm] {inputs/1101_1.jpg}};
\draw (0.5,4.5) rectangle (2.5,7.5);
\end {tikzpicture}
};

\node at (56,8) {$\mathtt{br} \to \mathtt{ebr}$};

\end {tikzpicture}
\end {center}
\end {minipage}}
\end {center}
\caption {Comparison over encrypted data with homomorphic encryption}
\label{fig:hom}
\end {figure}

Homomorphic encryption schemes \cite{DBLP:conf/stoc/Gentry09} makes it possible to compute certain functions over encrypted data.
For example, if $Enc$ is a homomorphic encryption scheme for multiplication then there is an operation $\otimes $ such that:
$$
c_1 = Enc (m_1) \wedge c_2 = Enc (m_2) \Rightarrow c_1 \otimes c_2 = Enc (m_1 \times m_2).
$$
Figure~\ref{fig:hom} presents an architecture $A_\mathsf{hom}$ derived from $A_\mathsf{hsm}$ in which the server performs the whole matching computation over encrypted data.
The user supplies a template that is sent encrypted to the server (denoted $\mathtt{ebs}$).
The server also owns an encrypted reference template $\mathtt{ebr}$.
The comparison, i.e. the computation of the distance between the templates, is done by the server, leading to the encrypted distance $\mathtt{edec}$, but the server does not get access to the biometric data or to the result.
This is made possible through the use a homomorphic encryption scheme.
On the other hand, the module gets the result, but does not get access to the templates.
Let us note that $A_\mathsf{hom}$ is just one of the possible ways to use homomorphic encryption in this context: the homomorphic computation of the distance could actually be made by another component (for example the terminal itself) since it does not lead to any leak of biometric data.

The homomorphic property of the encryption scheme needed for this application depends on the matching algorithm.
An option is to resort to a fully homomorphic encryption scheme (FHE) \cite{DBLP:conf/stoc/Gentry09} as in the solution described in \cite{DBLP:conf/icip/Troncoso-PastorizaP12} which uses a variant of a FHE scheme for face-recognition.
However, schemes with simpler homomorphic functionalities can also be sufficient
(examples can be found in \cite{DBLP:conf/acisp/BringerCIPTZ07,DBLP:conf/esorics/BlantonG11}).
Since we describe our solutions at the architecture level, we do not need to enter into details regarding the chosen homomorphic scheme.
We just need to assume the existence of a homomorphic matching function $Hom\text{-}\mu$ with the following properties captured by the algorithmic knowledge relations:
\begin {multline}
\{
\mathtt{ebr} = Enc (\mathtt{br}),
\mathtt{ebs} = Enc (\mathtt{bs}),
\\
\mathtt{edec} = Hom\text{-}\mu (\mathtt{ebr}, \mathtt{ebs}, \mathtt{thr})
\} \triangleright
\{Dec (\mathtt{edec}) = \mu (\mathtt{br}, \mathtt{bs}, \mathtt{thr})\}
\label{eq:homprop}
\end {multline}
The dependence relations include the following:
$\{(\mathtt{bs}, \{\mathtt{rd}\})$,
$(\mathtt{ebs}, \{\mathtt{bs}\}) \} \subseteq Dep_\mathsf{T}$;
$(\mathtt{ebr}, \{\mathtt{br}\}) \in Dep_\mathsf{I}$;
$\{(\mathtt{br}, \{\mathtt{ebr}\})$,
$(\mathtt{bs}, \{\mathtt{ebs}\})$,
$(\mathtt{dec}, \{\mathtt{edec}\})\} \subseteq Dep_\mathsf{M}$.
Architecture $A_\mathsf{hom}$ is defined as follows:
\begin {align*}
A_\mathsf{hom}
&
:= \big\{
Has_\mathsf{I} (\mathtt{br}), Has_\mathsf{U} (\mathtt{rd}), Has_\mathsf{S} (\mathtt{thr}),
Compute_\mathsf{I} (\mathtt{ebr} = Enc (\mathtt{br})),
\\
&
Receive_{\mathsf{S}, \mathsf{I}} (\{Attest_\mathsf{I} (\{\mathtt{ebr} = Enc (\mathtt{br})\})\}, \{\mathtt{ebr}\}),
Receive_{\mathsf{T}, \mathsf{U}} (\{\}, \{\mathtt{rd}\}),
\\
&
Compute_\mathsf{T} (\mathtt{bs} = Extract (\mathtt{rd})),
Compute_\mathsf{T} (\mathtt{ebs} = Enc (\mathtt{bs})),
\\
&
Receive_{\mathsf{S}, \mathsf{T}} (\{\}, \{\mathtt{ebs}\}),
Compute_\mathsf{S} (\mathtt{edec} = Hom\text{-}\mu (\mathtt{ebr}, \mathtt{ebs}, \mathtt{thr})),
\\
&
Receive_{\mathsf{T}, \mathsf{S}} (\mathcal{A}, \{\mathtt{edec}\}),
Verify_\mathsf{T} (Attest_\mathsf{I} (\{\mathtt{ebr} = Enc (\mathtt{br})\})),
\\
&
Verify_\mathsf{T} (Attest_\mathsf{S} (\{\mathtt{edec} = Hom\text{-}\mu (\mathtt{ebr}, \mathtt{ebs}, \mathtt{thr})\})),
Trust_{\mathsf{T}, \mathsf{S}},
\\
&
Trust_{\mathsf{T}, \mathsf{I}},
Receive_{\mathsf{M}, \mathsf{T}} (\{\}, \{\mathtt{edec}\}),
Compute_\mathsf{M} (\mathtt{dec} = Dec (\mathtt{edec})),
\\
&
Receive_{\mathsf{T}, \mathsf{M}} (\{Attest_\mathsf{M} (\{\mathtt{dec} = Dec (\mathtt{edec})\})\}, \{\mathtt{dec}\}),
Trust_{\mathsf{T}, \mathsf{M}},
\\
&
Verify_\mathsf{T} (Attest_\mathsf{M} (\{\mathtt{dec} = Dec (\mathtt{edec})\})),
Receive_{\mathsf{U}, \mathsf{T}} (\{\}, \{\mathtt{dec}\})
\big\}
\end {align*}
where the set $\mathcal{A}$ of attestations received by the terminal from the server is:
$\mathcal{A} := \{Attest_\mathsf{I} (\{\mathtt{ebr} = Enc (\mathtt{br})\}), Attest_\mathsf{S} (\{\mathtt{edec} = Hom\text{-}\mu (\mathtt{ebr}, \mathtt{ebs}, \mathtt{thr})\})\}$.

In order to prove that the terminal can establish the integrity of the result $\mathtt{dec}$, we can proceed in two steps, proving first the correctness of $\mathtt{edec}$ and then deriving the correctness of $\mathtt{edec}$ using the properties of homomorphic encryption.
The first step relies on the capacities of component $\mathsf{T}$ and the trust assumptions on components $\mathsf{I}$ and $\mathsf{S}$ using rules \textsf{K1} and \textsf{K5} respectively.

\begin {center}
\AxiomC {$Compute_\mathsf{T} (\mathtt{ebs} = Enc (\mathtt{bs})) \in A_\mathsf{hom}$}
\LeftLabel {\textsf{K1}}
\UnaryInfC {$A_\mathsf{hom} \vdash K_\mathsf{T} (\mathtt{ebs} = Enc (\mathtt{bs}))$}
\DisplayProof

\smallskip

\AxiomC {$Verify_\mathsf{T} (\{Attest_\mathsf{I} (\mathtt{ebr} = Enc (\mathtt{br}))\}) \in A_\mathsf{hom}$}
\AxiomC {$Trust_{\mathsf{T}, \mathsf{I}} \in A_\mathsf{hom}$}
\LeftLabel {\textsf{K5}}
\BinaryInfC {$A_\mathsf{hom} \vdash K_\mathsf{T} (\mathtt{ebr} = Enc (\mathtt{br}))$}
\DisplayProof

\smallskip

\AxiomC {$Verify_\mathsf{T} (\{Attest_\mathsf{S} (\mathtt{edec} = Hom\text{-}\mu (\mathtt{br}, \mathtt{bs}, \mathtt{thr}))\})$,
\
$Trust_{\mathsf{T}, \mathsf{S}} \in A_\mathsf{hom}$}
\LeftLabel {\textsf{K5}}
\UnaryInfC {$A_\mathsf{hom} \vdash K_\mathsf{T} (\mathtt{edec} = Hom\text{-}\mu (\mathtt{br}, \mathtt{bs}, \mathtt{thr}))$}
\DisplayProof
\end {center}
The second step can be done through the application of the deductive algorithmic knowledge regarding the homomorphic encryption property (with $LHS_1$ the left hand-side of equation (\ref{eq:homprop})) :

\begin {center}
\AxiomC {$LHS_1 \triangleright \{Dec (\mathtt{edec}) = \mu (\mathtt{br}, \mathtt{bs}, \mathtt{thr})\}$}
\AxiomC {$\forall Eq \in LHS_1: A_\mathsf{hom} \vdash K_\mathsf{T} (Eq)$}
\LeftLabel {\textsf{K$\triangleright$}}
\BinaryInfC {$A_\mathsf{hom} \vdash K_\mathsf{T} (Dec (\mathtt{edec}) = \mu (\mathtt{br}, \mathtt{bs}, \mathtt{thr}))$}
\DisplayProof
\end {center}
The desired property is obtained through the application of rules \textsf{K5} and \textsf{K$\triangleright$} exploiting the trust relation between $ \mathsf{T}$ and $\mathsf{M} $ and the transitivity of equality.

\begin {center}
\AxiomC {$Verify_\mathsf{T} (\{Attest_\mathsf{M} (\mathtt{dec} = Dec (\mathtt{edec}))\}) \in A_\mathsf{hom}$}
\AxiomC {$Trust_{\mathsf{T}, \mathsf{M}} \in A_\mathsf{hom}$}
\LeftLabel {\textsf{K5}}
\BinaryInfC {$A_\mathsf{hom} \vdash K_\mathsf{T} (\mathtt{dec} = Dec (\mathtt{edec}))$}
\DisplayProof
\end {center}

\begin {center}
\AxiomC {$A_\mathsf{hom} \vdash K_\mathsf{T} (Dec (\mathtt{edec}) = \mu (\mathtt{br}, \mathtt{bs}, \mathtt{thr}))$
\quad
$A_\mathsf{hom} \vdash K_\mathsf{T} (\mathtt{dec} = Dec (\mathtt{edec}))$}
\LeftLabel {\textsf{K$\triangleright$}}
\UnaryInfC {$A_\mathsf{hom} \vdash K_\mathsf{T} (\mathtt{dec} = \mu (\mathtt{br}, \mathtt{bs}, \mathtt{thr}))$}
\DisplayProof
\end {center}
As far as privacy is concerned, the main property that $A_\mathsf{hom}$ is meant to ensure is that no component (except the issuer) has access to the biometric references.
Rule \textsf{HN} makes it possible to prove that \textsf{U}, \textsf{T}, and \textsf{S} never get access to \texttt{br}, as in Section~\ref{s:proctemp}.
The same rule can be applied here to prove $A_\mathsf{hom} \nvdash Has_{\mathsf{M}} (\mathtt{ebr})$ exploiting the fact that neither $(\mathtt{br}, \{\mathtt{edec}\})$ nor $(\mathtt{br}, \{\mathtt{dec}\})$ belong to $Dep_{\mathsf{M}}$.

\subsection {The Match-On-Card technology}
\label{s:moc}

\begin {figure} [ht]
\begin {center}
\fbox {\begin {minipage}{.98\textwidth} \footnotesize
\begin {center}
\begin {tikzpicture}[scale=0.1]

\node [above] at (-35,-15) {$\mathsf{U}$};
\node [above] at ( 0,-15) {$\mathsf{T}$};
\node [above] at ( 43,-15) {$\mathsf{C}$};

\node [scale=1.2] at (-35,0)
{
\begin {tikzpicture} [scale=0.1]
\draw (0,10) circle (2);
\draw (0,8) -- (0,4);
\draw (0,4) -- (-2,0);
\draw (0,4) -- (2,0);
\draw (0,6) -- (-2,6);
\draw (0,6) -- (2,6);
\end {tikzpicture}
};

\draw [->] (-30,2) -- (-10,2);
\node [above] at (-20,2) {$\mathtt{rd}$};
\draw [<-] (-30,-2) -- (-10,-2);
\node [below] at (-20,-2) {$\mathtt{dec}$};

\node [scale=0.4] at (0,0)
{
\begin {tikzpicture} [scale=0.5]
\draw [rounded corners] (0,0) rectangle (8,10);
\node at (6.5,4) {\includegraphics [width=1.1cm] {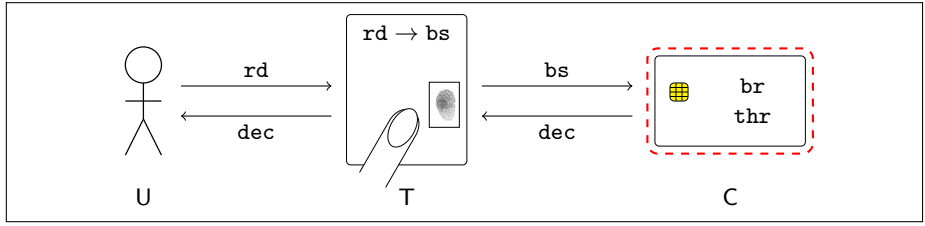}};
\draw (5.5,2.5) rectangle (7.5,5.5);
\draw [rotate=-30,fill=white] (1,0) -- (1,4) arc (180:0:1 and 1.3) -- (3,0);
\draw [rotate=-30,fill=white] (2,4) ellipse (0.8 and 1.4);
\end {tikzpicture}
};

\node [above] at (0,7) {$\mathtt{rd} \to \mathtt{bs}$};

\draw [->] (10,2) -- (30,2);
\node [above] at (20,2) {$\mathtt{bs}$};
\draw [<-] (10,-2) -- (30,-2);
\node [below] at (20,-2) {$\mathtt{dec}$};

\node [scale=0.4] at (43,0)
{
\begin {tikzpicture} [scale=0.5]
\draw [rounded corners] (0,0) rectangle (10,6);
\draw [fill=yellow, rounded corners] (1,3) rectangle (2.2,4.2);
\draw (1.4,3) -- (1.4,4.2);
\draw (1.8,3) -- (1.8,4.2);
\draw (1,3.3) -- (2.2,3.3);
\draw (1,3.6) -- (2.2,3.6);
\draw (1,3.9) -- (2.2,3.9);
\end {tikzpicture}
};

\node at (46,2) {$\mathtt{br}$};
\node at (46,-2) {$\mathtt{thr}$};
\draw [rounded corners, red, dashed,thick] (32,-7) rectangle (54,7);

\end {tikzpicture}
\end {center}
\end {minipage}}
\end {center}
\caption {Biometric verification using the Match-On-Card technology}
\label{fig:moc}
\end {figure}

Another solution can be considered when the purpose of the system is authentication rather than identification.
In this case, it is not necessary to store a database of biometric reference templates and a (usually unique) reference template can be stored on a smart card.
A smart card based privacy preserving architecture has been proposed recently which relies on the idea of using the card not only to store the reference template but also to perform the matching itself.
Since the comparison is done inside the card the reference template never leaves the card.
In this \textit{Match-On-Card} (MOC) technology~\cite{ISOIEC24787,NISTMOC,GB07} (also called \textit{com\-pa\-rison-on-card}), the smart card receives the fresh biometric template, carries out the comparison with its reference template, and sends the decision back (as illustrated in Figure~\ref{fig:moc}).

In this architecture, the terminal is assumed to trust the smart card.
This trust assumption is justified by the fact that the card is a tamper-resistant hardware element.
This architecture is simpler than the previous ones but not always possible in practice (for a combination of technical and economic reasons) and may represent a shift in terms of trust if the smart card is under the control of the user.

More formally, the MOC architecture is composed of a user \textsf{U}, a terminal \textsf{T}, and a card \textsf{C}.
The card \textsf{C} attests that the templates $\mathtt{br}$ and $\mathtt{bs}$ are close (with respect to the threshold $\mathtt{thr}$):
\begin {align*}
A_\mathsf{moc} := \big\{
&
Has_\mathsf{C} (\mathtt{br}), Has_\mathsf{U} (\mathtt{rd}), Has_\mathsf{C} (\mathtt{thr}),
Receive_{\mathsf{T}, \mathsf{U}} (\{\}, \{\mathtt{rd}\}),
\\
&
Compute_\mathsf{T} (\mathtt{bs} = Extract (\mathtt{rd})),
Receive_{\mathsf{C}, \mathsf{T}} (\{\}, \{\mathtt{bs}\}),
\\
&
Compute_\mathsf{C} (\mathtt{dec} = \mu (\mathtt{br}, \mathtt{bs}, \mathtt{thr})),
Receive_{\mathsf{U}, \mathsf{T}} (\{\}, \{\mathtt{dec}\}),
\\
&
Receive_{\mathsf{T}, \mathsf{C}} (\{Attest_\mathsf{C} (\mathtt{dec} = \mu (\mathtt{br}, \mathtt{bs}, \mathtt{thr}))\}, \{\mathtt{dec}\}),
\\
&
Verify_\mathsf{T} (\{Attest_\mathsf{C} (\mathtt{dec} = \mu (\mathtt{br}, \mathtt{bs}, \mathtt{thr}))\}),
Trust_{\mathsf{T}, \mathsf{C}}
\big\}
\end {align*}
Using rule \textsf{HN}, it is easy to show that no component apart from $\mathsf{C}$ gets access to~$\mathtt{br}$.
The proof of the integrity property relies on the capacities of component $\mathsf{T}$ and the trust assumption on component $\mathsf{C}$ using rules \textsf{K1} and \textsf{K5} respectively.

\section{Extension of the framework to information leakage}
\label{s:extension}
\subsection {Extension of the architecture language}
\label{s:extarch}

Motivated by the need to analyse the inherent leakage of the result of a matching between two biometric data in biometric systems (cf. \cite{DBLP:conf/iih-msp/BringerCS10,DBLP:journals/tifs/SimoensBCS12,DBLP:conf/indocrypt/PagninDAM14}), we now propose an extension of the formal framework sketched in Section~\ref{s:framework}, in which the information leaking through several executions can be expressed.

We highlights the difference with the framework introduced in Section~\ref{s:framework} without repeating their common part.
The term language we use is now the following.
$$
\begin {array} {ccccccc}
T & ::= & \tilde{X} & | & c & | & F (\tilde{X}_1, \dots, \tilde{X}_m, c_1, \dots, c_q) \\
\tilde{X} & ::= & X & | & X[k] \\
\end {array}
$$
Functions may take as parameters both variables and constants.
Variables $\tilde{X}$ can be simple variables or arrays of variables.
If $X$ is an array, $Range (X)$ denotes its size.

In this extended framework, in addition to defining a set of primitives, an architecture can also provide a bound on the number of times a primitive can be used.
$$
\begin {array} {l}
\begin {array} {rrlclclcl}
\makebox[7mm]{\hfill$A$} & ::= &
\multicolumn{5}{l}{\{R\}} \\
R & ::= & Has_i^{(n)} (X) & | & Has_i (c) & | & \multicolumn{3}{l} {Receive_{i, j}^{(n)} (\{St\}, \{X\} \cup \{c\})} \\
& | & Trust_{i, j} & | & Reset & | & Compute_G^{(n)} (X = T) & | & Verify^{(n)}_i (\{St\}) \\
\end {array} \\ \\
\begin {array} {rrlclcccrrl}
\makebox[7mm]{\hfill$St$} & ::= & Pro & | & Att & & & & Att & ::= & Attest_i (\{Eq\}) \\
Pro & ::= & \multicolumn{6}{l}{Proof_i (\{P\}) \qquad} & Eq &::= & Pred (T_1, \dots, T_m) \\
P & ::= & Att & | & Eq \\
\end {array}
\end {array}
$$
The superscript notation ${}^{(n)}$ denotes that a primitive can be carried out at most $n \in (\mathbb{N} \setminus \{0\}) \cup \{\infty\}$ times by the component(s) -- where ($\forall n' \in \mathbb{N}$: $n' < \infty$).
We assume that $n$ is never equal to 0.
$\mathsf{mul} (\alpha)$ denotes the multiplicity $(n)$ of the primitive $\alpha$, if any.
The $Reset$ primitive is used to reinitialize the whole system.

As in the initial model, consistency assumptions are made about the architectures to avoid meaningless definitions. 
For instance, we require that components carry out computations only on the values that they have access to (either through $Has$, $Compute$, or $Receive$).
We also require that all multiplicities $n$ specified by the primitives are identical in a consistent architecture.
As a result, a consistent architecture $A$ is parametrized by an integer $n \geq 1$ (we note $A (n)$ when we want to make this integer explicit).

A key concept for the definition of the semantics is the notion of trace. A trace is a sequence of events and an event\footnote{Except for the \textit{Session} event.} is an instantiation of an architectural primitive\footnote{Except for \textit{Trust} primitives, which cannot be instantiated into events because they are global assumptions.}.
The notion of successive sessions is caught by the addition of a $Session$ event\footnote{Computations can involve different values of the same variables from different sessions.} .
A trace $\theta$ of events is said {compatible} with a consistent architecture $A (n)$ if all events in $\theta$ (except the computations) can be obtained by instantiation of some architectural primitive from~$A$, and if the number of events between two $Reset$ events corresponding to a given primitive is less than the bound $n$ specified by the architecture.
We denote by $T (A)$ the set of traces which are compatible with an architecture $A$.

$$
\begin {array} {rrlclclcl}
\theta & ::= & \mathsf{Seq} (\epsilon) \\
\epsilon & ::= & Has_i (X:V) & | & Has_i (c) & | & \multicolumn{3}{l} {Receive_{i, j} (\{St\}, \{X:V\} \cup \{c\})} \\
& | & Session & | & Reset & | & Compute_G (X = T) & | & Verify_i (\{St\}) \\
\end {array}
$$
An event can instantiate variables $X$ with specific values $V$.
Constants always map to the same value.
Let $Val$ be the set of values the variables and constants can take.
The set $Val_\bot$ is defined as $Val \cup \{\bot\}$ where $\bot \not\in Val$ is a specific symbol used to denote that a variable or a constant has not been assigned yet.

The semantics of an architecture follows the approach introduced in~\cite{DBLP:conf/stm/AntignacM14}.
Each component is associated with a state.
Each event in a trace of events affects the state of each component involved by the event.
The semantics of an architecture is defined as the set of states reachable by compatible traces.

The state of a component is either the $Error$ state or a pair consisting of:
(i) a variable state assigning values to variables, and (ii) a property state defining what is known by a component.
$$
\begin {array} {rcl}
State_\bot & = & (State_V \times State_P) \cup \{Error\} \\
State_V & = & Var \cup Const \to \mathsf{List} (Val_\bot) \\
State_P & = & \{Eq\} \cup \{Trust_{i, j}\}
\end {array}
$$
The data structure $\mathsf{List}$ over a set $S$ denotes the finite ordered lists of elements of $S$,
$\mathsf{size} (L)$ denotes the size of the list $L$, and
$()$ is the empty list.
For a non-empty list $L = (e_1, \dots, e_n) \in S^n$ where $\mathsf{size} (L) = n \geq 1$, $L[m]$ denotes the element $e_m$ for $1 \leq m \leq n$, $\mathsf{last} (L)$ denotes $L [n]$, and $\mathsf{append} (L, e)$ denotes the list $(e_1, \dots, e_n, e) \in S^{n + 1}$.
Let $\sigma := (\sigma_1, \dots, \sigma_N)$ denote the global state (\textit{i.e.} the list of states of all components) defined over $(State_\bot)^N$ and
$\sigma_i^v$ and $\sigma_i^{pk}$ denote, respectively, the variable and the knowledge state of the component $C_i$.

The variable state assigns values to variables and to constants (each constant is either undefined or taking a single value).
$\sigma_i^v (X) [m]$ (resp. $\sigma_i^v (c) [m]$) denotes the $m$-th entry of the variable state of $X \in Var$ (resp. $c \in Const$).
The initial state of an architecture $A$ is denoted by $Init^A = \langle Init^A_1, \dots, Init^A_{N} \rangle$ where:
$\forall C_i$: $Init^A_i =$ ($Empty$, \{$Trust_{i, j} \ | \ \exists C_j$: $Trust_{i, j} \in A$\}).
$Empty$ associates to each variable and constant a list made of a single undefined value $(\bot)$.
We assume that, in the initial state, the system is in its first session.
Alternatively, we could set empty lists in the initial state and assume that every consistent trace begins with a $Session$ event.

Let $S_T: Trace \times (State_\bot)^{N} \to (State_\bot)^{N}$ and $S_E: Event \times (State_\bot)^{N} \to (State_\bot)^{N}$ be the following two functions.
$S_T$ is defined recursively by iteration of $S_E$:
for all state $\sigma \in (State_\bot)^{N}$, event $\epsilon \in Event$ and consistent trace $\theta \in Trace$, $S_T (\langle \rangle, \sigma) = \sigma$ and $S_T (\epsilon \cdot \theta, \sigma) = S_T (\theta, S_E (\epsilon, \sigma))$.
The modification of a state is noted $\sigma [\sigma_i/(v, pk)]$ the variable and knowledge states of $C_i$ are replaced by $v$ and $pk$ respectively.
$\sigma[\sigma_i /Error]$ denotes that the $Error$ state is reached for component $C_i$.
We assume that a component reaching an $Error$ state no longer gets  involved in any later action (until a reset of the system).
The function $S_E$ is defined event per event.

The effect of $Has_i (X:V)$ and $Receive_{i,j} (S, \{(X:V)\})$ on the variable state of component $C_i$ is the replacement of the last value of the variable $X$ by the value $V$:
$\mathsf{last} (\sigma_i^v (X)) := V$.
This effect is denoted by $\sigma_i^v [X / V]$:
$$
S_E (Has_i (X:V), \sigma) =
S_E (Receive_{i, j} (S, \{X:V\}), \sigma) =
\sigma[\sigma_i / (\sigma_i^v[X/V], \sigma^{pk}_i)].
$$
In the case of constants, the value $V$ is determined by the interpretation of $c$ (as in the function symbols in the computation).

The effect of $Compute_G (X = T)$ is to assign to $X$, for each component $C_i \in G$, the value $V$ produces by the evaluation (denoted $\varepsilon$) of $T$.
The new knowledge is the equation $X = T$.
A computation may involve values of variables from different sessions.
As a result, some consistency conditions must be met, otherwise an error state is reached:
$$
S_E (Compute_G (X = T), \sigma) =
\begin {cases} \sigma[\forall C_i \in G: \sigma_i / (\sigma_i^v[X / V], \sigma^{pk}_i \cup \{X = T\})] \\
\qquad \text{if the condition on the computation holds,} \\
\sigma[\sigma_i / Error] \ \text{otherwise,}
\end {cases}
$$
where $V := \varepsilon \left(T, \cup_{C_i \in G} \sigma_i^v\right)$.
For each $\tilde{X}^{(n)} \in T$, the evaluation of $T$ is done with respect to the $n$ last values of $\tilde{X}$ that are fully defined.
An error state is reached if $n$ such values are not available.
The condition on the computation is then:
$\forall C_i \in G, \tilde{X}^{(n)} \in T$: $\mathsf{size} \big(\big\{m \ \big| \ \sigma_i^v \big(V \big(\tilde{X}\big)\big) [m] \text{ is fully defined}\big\}\big) \geq n$.

Semantics of the verification events are defined according to the (implicit) semantics of the underlying verification procedures.
In each case, the knowledge state of the component is updated if the verification passes, otherwise the component reaches an $Error$ state.
The variable state is not affected.
\begin {align*}
S_E (Verify_i (Proof_j (E)), \sigma) & = \begin {cases}
\sigma[\sigma_i/ (\sigma_i^v, \sigma_i^{pk} \cup new^{pk}_{Proof})] \\ \qquad \text{if the proof is valid}, \\
\sigma[\sigma_i / Error] \ \text{otherwise},
\end {cases} \\
S_E (Verify_i (Attest_j (E)), \sigma) & = \begin {cases}
\sigma[\sigma_i / (\sigma_i^v, \sigma_i^{pk} \cup new^{pk}_{Attest})] \\ \qquad \text{if the attestation is valid}, \\
\sigma[\sigma_i / Error] \ \text{otherwise}.
\end {cases}
\end {align*}
The new knowledge $new^{pk}_{Proof}$ and $new^{pk}_{Attest}$ are defined as:
\begin {align*}
new^{pk}_{Proof} & := \left\{Eq \ \middle| \ Eq \in E \ \vee \ \left(\begin {array} {l}
\exists C_k: Attest_k (E') \in E \ \wedge \ Eq \in E' \\
\wedge \ Trust_{i, k} \in \sigma_i^{pk}
\end {array}\right)\right\} \text{ and} \\
new^{pk}_{Attest} & := \{Eq \ | \ Eq \in E \ \wedge \ Trust_{i, j} \in \sigma_i^{pk}\}.
\end {align*}
In the session case, the knowledge state is reinitialized and a new entry is added in the variable states:
$$
S_E (Session, \sigma) = \sigma [\forall i:\sigma_i / (upd^v, \{Trust_{i, j} \ | \ \exists C_j: Trust_{i, j} \in A\})],
$$
where the new variable state $upd^v$ is such that $\sigma_i^v (X) := \mathsf{append} (\sigma_i^v (X), \bot)$ for all variables $X \in Var$, and $\sigma_i^v (c) := \mathsf{append} (\sigma_i^v (c), \mathsf{last} (\sigma_i^v (c)))$ for all constants $c \in Const$.
The session event is not local to a component, all component states are updated.
As a result, we associate to each global state $\sigma$ a unique number, noted $\mathsf{s} (\sigma)$, which indicates the number of sessions.
In the initial state, $\mathsf{s} (\sigma) := 1$, and at each $Session$ event, $\mathsf{s} (\sigma)$ is incremented.

In the reset case, all values are dropped and the initial state is restored:
$S_E (Reset, \sigma) = Init^{A}$.

This ends the definition of the semantics of trace of events.
The semantics $S (A)$ of an architecture $A$ is defined as the set of states reachable by compatible traces.

\subsection {Extension of the privacy logic}
\label{s:extlog}

The privacy logic is enhanced to express access to $n$ values of a given variable.
The formula $Has_i$ represents $n \geq 1$ accesses by $C_i$ to some variable $X$.
$$
\begin {array} {rclclclclclcl}
\varphi & ::= & Has_i (X^{(n)}) & | & Has_i (c) & | & Has^{none}_i (X) & | & Has^{none}_i (c) & | & K_i (Eq) & | & \varphi_1 \wedge \varphi_2 \\
Eq & ::= & \multicolumn{5}{l}{Pred (T_1, \dots, T_m)}
\end {array}
$$
Several values of the same variables from different sessions can provide information about other variables, which is expressed through the dependence relation.

The {semantics $S (\varphi)$ of a property $\varphi \in \mathcal{L}_P$} remains defined as the set of architectures where $\varphi$ is satisfied.
The fact that $\varphi$ is satisfied by a (consistent) architecture $A$ is defined as follows.

\begin {itemize}
\item $A$ satisfies $Has_i (X^{(n)})$
if there is a reachable state in which $X$ is fully defined (at least) $n \geq 1$ times.



\item $A$ satisfies $Has_i (c)$
if there is a reachable state in which $c$ is fully defined.



\item $A$ satisfies $Has^{none}_i (X)$ (resp. $Has^{none}_i (c)$) if no compatible trace leads to a state in which $C_i$ assigns a value to $X$ (resp. $c$).




\item $A$ satisfies $K_i (Eq)$ if for all reachable states, there exists a state in the same session in which $C_i$ can derive $Eq$.



\item $A$ satisfies $\varphi_1 \wedge \varphi_2$ if $A$ satisfies $\varphi_1$ and $A$ satisfies $\varphi_2$.

\end {itemize}

\begin {figure} [t!]
\begin {center} \small
\fbox { \begin {minipage} {.97\textwidth}
\AxiomC {$Has_i^{(n)} (X) \in A$}
\LeftLabel {\textsf{H1}}
\UnaryInfC {$A \vdash Has_i (X^{(n)})$}
\DisplayProof
\hfill
\AxiomC {$Receive_{i, j}^{(n)} (S, E) \in A$}
\AxiomC {$X \in E$}
\LeftLabel {\textsf{H2}}
\BinaryInfC {$A \vdash Has_i (X^{(n)})$}
\DisplayProof
\hfill
\AxiomC {$A \nvdash Has_i (X^{(1)})$}
\LeftLabel {\textsf{HN}}
\UnaryInfC {$A \vdash Has^{none}_i (X)$}
\DisplayProof

\smallskip

\AxiomC {$Has_i (c) \in A$}
\LeftLabel {\textsf{H1'}}
\UnaryInfC {$A \vdash Has_i (c)$}
\DisplayProof
\hfill
\AxiomC {$Receive_{i, j}^{(n)} (S, E) \in A$}
\AxiomC {$c \in E$}
\LeftLabel {\textsf{H2'}}
\BinaryInfC {$A \vdash Has_i (c)$}
\DisplayProof
\hfill
\AxiomC {$A \nvdash Has_i (c)$}
\LeftLabel {\textsf{HN'}}
\UnaryInfC {$A \vdash Has^{none}_i (c)$}
\DisplayProof

\smallskip

\AxiomC {$Compute_G^{(n)} (X = T) \in A$}
\AxiomC {$C_i \in G$}
\LeftLabel {\textsf{H3}}
\BinaryInfC {$A \vdash Has_i (X^{(n)})$}
\DisplayProof
\hfill
\AxiomC {$A \vdash Has_i (X^{(n)})$}
\AxiomC {$1 \leq m \leq n$}
\LeftLabel {\textsf{H4}}
\BinaryInfC {$A \vdash Has_i (X^{(m)})$}
\DisplayProof

\smallskip

\AxiomC {$Dep_i (Y, \mathcal{X})$}
\AxiomC {$\forall X^{(n)} \in \mathcal{X}$: $A \vdash Has_i (X^{(n)})$}
\AxiomC {$\forall c \in \mathcal{X}$: $A \vdash Has_i (c)$}
\LeftLabel {\textsf{H5}}
\TrinaryInfC {$A \vdash Has_i (Y^{(1)})$}
\DisplayProof

\smallskip

\AxiomC {$Dep_i (c, \mathcal{X})$}
\AxiomC {$\forall X^{(n)} \in \mathcal{X}$: $A \vdash Has_i (X^{(n)})$}
\AxiomC {$\forall c' \in \mathcal{X}$: $A \vdash Has_i (c')$}
\LeftLabel {\textsf{H5'}}
\TrinaryInfC {$A \vdash Has_i (c)$}
\DisplayProof

\smallskip

\AxiomC {$Compute_G^{(n)} (X = T) \in A$}
\AxiomC {$C_i \in G$}
\LeftLabel {\textsf{K1}}
\BinaryInfC {$A \vdash K_i (X = T)$}
\DisplayProof
\hfill
\AxiomC {$A \vdash \varphi_1$}
\AxiomC {$A \vdash \varphi_2$}
\LeftLabel {\textsf{I$\wedge$}}
\BinaryInfC {$A \vdash \varphi_1 \wedge \varphi_2$}
\DisplayProof

\smallskip

\AxiomC {$E \triangleright_i Eq_0$}
\AxiomC {$\forall Eq \in E$: $A \vdash K_i (Eq)$}
\LeftLabel {\textsf{K$\triangleright$}}
\BinaryInfC {$A \vdash K_i (Eq_0)$}
\DisplayProof
\hfill
\AxiomC {$A \vdash K_i (Eq_1)$}
\AxiomC {$A \vdash K_i (Eq_2)$}
\LeftLabel {\textsf{K$\wedge$}}
\BinaryInfC {$A \vdash K_i (Eq_1 \wedge Eq_2)$}
\DisplayProof

\smallskip

\AxiomC {$Verify_i^{(n)} (Proof_j (E)) \in A$}
\AxiomC {$Eq \in E$}
\LeftLabel {\textsf{K3}}
\BinaryInfC {$A \vdash K_i (Eq)$}
\DisplayProof

\smallskip

\AxiomC {$Verify_i^{(n)} (Proof_j (E)) \in A$}
\AxiomC {$Attest_k (E') \in E$}
\AxiomC {$Eq \in E'$}
\AxiomC {$Trust_{i, k} \in A$}
\LeftLabel {\textsf{K4}}
\QuaternaryInfC {$A \vdash K_i (Eq)$}
\DisplayProof

\smallskip

\AxiomC {$Verify^{(n)}_i (Attest_j (E)) \in A$}
\AxiomC {$Trust_{i, j} \in A$}
\AxiomC {$Eq \in E$}
\LeftLabel {\textsf{K5}}
\TrinaryInfC {$A \vdash K_i (Eq)$}
\DisplayProof
\end {minipage}}
\end {center}
\caption {Set of deductive rules for the extended privacy logic}
\label{fig:axioms}
\end {figure}

A set of {deductive rules} for this privacy logic is given in Figure~\ref{fig:axioms}.
One can show that this axiomatics is sound and complete with respect to the semantics above.
The soundness theorem states that for all $A$, if $A \vdash \varphi$, then $A \in S (\varphi)$.
Completeness means that for all $A$, if $A \in S (\varphi)$ then $A \vdash \varphi$.

Due to the length of the proofs and the lack of place, we only give sketch for these proofs.
Soundness is proved by induction on the derivation tree.
For each theorem $A \vdash \varphi$, one can find traces satisfying the claimed property, or show that all traces satisfy the claimed property (depending on the kind of property).
Completeness is shown by induction on the property $\varphi$.
For each property belonging to the semantics, one can exhibit a tree that derives it from the architecture.

A trace is said to be a {covering trace} if it contains an event corresponding to each primitive specified in an architecture $A$ (except trust relations) and if for each primitive it contains as much events as the multiplicity $^{(n)}$ of the primitive.
As a first step to prove soundness, it is shown that for all consistent architecture $A$, there exists a consistent trace $\theta \in T (A)$ that covers $A$.

Then the soundness is shown by induction on the depth of the tree $A \vdash \varphi$.
\begin {itemize}

\item
Let us assume that $A \vdash Has_i (X^{(n)})$, and that the derivation tree is of depth 1.
By definition of $\mathcal{D}$, such a proof is obtained by application of (\textsf{H1}), (\textsf{H2}) or (\textsf{H3}).
In each case, it is shown (thanks to the existence of covering traces) that an appropriate trace can be found in the semantics of $A$, hence $A \in S (Has_i (X^{(n)}))$.
The case of $A \vdash Has_i (c)$ is very similar.

\item
Let us assume that $A \vdash K_i (Eq)$, and that the derivation tree is of depth~1.
By definition of $\mathcal{D}$, such a proof is obtained by application of (\textsf{K1}), (\textsf{K2}), (\textsf{K3}), (\textsf{K4}) or (\textsf{K5}).
In each case, starting from a state $\sigma' \in S_i (A)$ such that $\mathsf{s} (\sigma') \geq n$, it is first shown that there exists a covering trace $\theta \geq \theta'$ that extends $\theta'$ and that contains $n$ corresponding events $Compute_G (X = T) \in \theta$ in $n$ distinct sessions (for the \textsf{K1} case, and other events for the other rules).
Then by the properties of the deductive algorithmic knowledge, it is shown that the semantics of the property $A \in S (K_i (X = T))$ holds.

\item
Let us assume that $A \vdash Has_i (X^{(n)})$, and that the derivation tree is of depth strictly greater than 1.
By definition of $\mathcal{D}$, such a proof is obtained by application of (\textsf{H4}) or (\textsf{H5}).

In the first case, by the induction hypothesis and the semantics of properties, there exists a reachable state $\sigma \in S (A)$ and $n$ indices $i_1, \dots, i_n$ such that $\sigma_i^v (X) [i_l]$ is fully defined for all $l \in [1, n]$.
This gives, \textit{a fortiori}, $A \in S (Has_i (X^{(m)}))$ for all $m$ such that $1 \leq m \leq n$.

In the second case, we have that $(Y, \{X_1^{(n_1)}, \dots, X_m^{(n_m)}, c_1, \dots, c_q\}) \in Dep_i$, that $\forall l \in [1, m]:$ $A \vdash Has_i (X_l^{(n_l)})$ and $\forall l \in [1, q]:$ $A \vdash Has_i (c_l)$.
The proof shows the existence of a covering trace that contains an event $Compute_G$ ($Y$ $=$ $T$) (where $i \in G$), allowing to conclude that $A \in S (Has_i (Y^{(1)}))$.

\smallskip

Again, the corresponding cases for constant are very similar.

\item
A derivation for $Has^{none}$ is obtained by application of (\textsf{HN}).
The proof assume, towards a contradiction, that $A \not\in S (Has^{none}_i (X))$.
It is shown, by the architecture semantics, that there exists a compatible trace that enable to derive $A \vdash Has^{(1)}_i (X)$.
However, since (\textsf{HN}) was applied, we have $A \nvdash Has^{(1)}_i (X)$, hence a contradiction.

\item
The last case (the conjunction $\wedge$) is fairly straightforward.

\end {itemize}
The completeness is proved by induction over the definition of $\varphi$.
\begin {itemize}

\item
Let us assume that $A \in S (Has_i (X^{(n)}))$.
By the architecture semantics and the semantics of traces, it is shown that the corresponding traces either contain events where $X$ is computed, received or measured, or that some dependence relation on $X$ exists.
In the first case, we have $A \vdash Has_i (X^{(n)})$ by applying (respectively) (\textsf{H1}), (\textsf{H2}), or (\textsf{H3}) (after an eventual application of (\textsf{H4})).
In the last case, the proof shows how to exhibit a derivation tree to obtain $A \vdash Has_i (X^{(n)})$ (the (\textsf{H5}) rule is used).

\item
Let us assume that $A \in S (Has^{none}_i (X))$.
By the semantics of properties, this means that in all reachable states, $X$ does not receive any value.
The proof shows that $A \nvdash S (Has_i (X^{(1)}))$, otherwise $A \in S (Has^{none}_i (X))$ would be contradicted.
So as a conclusion, $A \vdash Has^{none}_i (X)$ by applying (\textsf{HN}).

\item
The constant cases $A \in S (Has_i (c)$ and $A \in S (Has^{none}_i (c))$ case are similar to the variable cases.

\item
Let us assume that $A \in S (K_i (Eq))$.
By the semantics of properties this means that for all reachable states, there exists a later state in the same session where the knowledge state enables to derive $Eq$.
By the semantics of architecture, we can exhibit a compatible trace that reaches a state where $Eq$ can be derived.
By the semantics of compatible traces, the proof shows, by reasoning on the events on the traces, that $A \vdash K_i (Eq)$ by applying either (\textsf{K1}), (\textsf{K2}), (\textsf{K3}), (\textsf{K4}) or (\textsf{K5}).

\item
Finally the conjunctive case is straightforward.
\end {itemize}

\section {Extension of the Match-On-Card to the identification paradigm}
\label{s:mocindent}

We now show of the extended framework can be used to reason about the privacy properties of a biometric system where some information leaks after several sessions of the same protocol.

The biometric system introduced in~\cite{DBLP:conf/cost/BringerCKK09} aims at extending the MOC technology (cf. Section~\ref{s:moc}) to the identification paradigm.
A {quantized} version -- corresponding to short binary representations of the templates -- of the database is stored inside a secure module, playing the role of the card in the MOC case.
From each biometric reference template, a {quantization} is computed, using typically a secure sketch scheme~\cite{DBLP:conf/ccs/JuelsW99,DBLP:conf/eurocrypt/DodisRS04}.
The reference database is encrypted and stored outside the secure module, whereas the quantizations of the templates are stored inside.

\begin {figure} [ht]
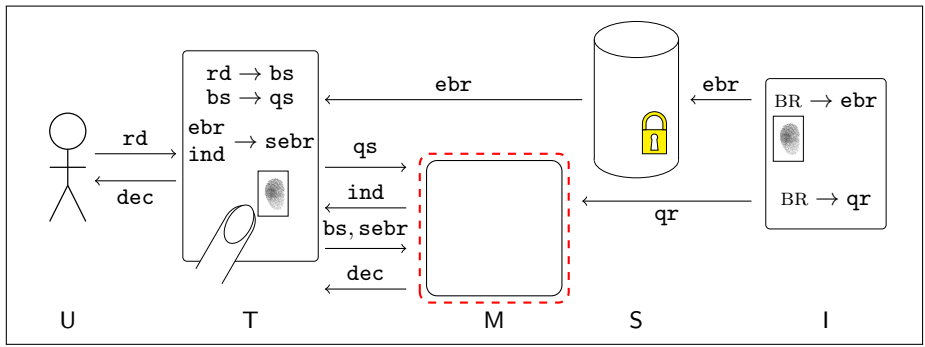

\begin {center}
\fbox {\begin {minipage}{.98\textwidth} \small
\begin {center}
\begin {tikzpicture}[scale=0.09]

\node [above] at (-56,-25) {$\mathsf{U}$};
\node [scale=1.2] at (-56,0)
{
\begin {tikzpicture} [scale=0.1]
\draw (0,10) circle (2);
\draw (0,8) -- (0,4);
\draw (0,4) -- (-2,0);
\draw (0,4) -- (2,0);
\draw (0,6) -- (-2,6);
\draw (0,6) -- (2,6);
\end {tikzpicture}
};

\draw [->] (-52,2) -- (-40,2);
\node [above] at (-46,2) {$\mathtt{rd}$};
\draw [<-] (-52,-2) -- (-40,-2);
\node [below] at (-46,-2) {$\mathtt{dec}$};

\node [above] at (-29,-25) {$\mathsf{T}$};
\node [scale=0.4] at (-29,0)
{
\begin {tikzpicture} [scale=0.5]
\node at (6,4.5) {\includegraphics [width=1.1cm] {inputs/1101_1.jpg}};
\draw [rounded corners] (0,0) rectangle (9,14);
\draw (5,3) rectangle (7,6);
\draw [rotate=-30,fill=white] (1,0) -- (1,4) arc (180:0:1 and 1.3) -- (3,0);
\draw [rotate=-30,fill=white] (2,4) ellipse (0.8 and 1.4);
\end {tikzpicture}
};

\node at (-29,14) {$\mathtt{rd} \to \mathtt{bs}$};
\node at (-29,10) {$\mathtt{bs} \to \mathtt{qs}$};
\node at (-35.5,6) {$\mathtt{ebr}$};
\node at (-25.5,4) {$\to \mathtt{sebr}$};
\node at (-35.5,2) {$\mathtt{ind}$};

\draw [->] (-18,-0) -- (-6,0);
\node [above] at (-12,0) {$\mathtt{qs}$};
\draw [<-] (-18,-6) -- (-6,-6);
\node [above] at (-12,-6) {$\mathtt{ind}$};
\draw [->] (-18,-12) -- (-6,-12);
\node [above] at (-12,-12) {$\mathtt{bs}, \mathtt{sebr}$};
\draw [<-] (-18,-18) -- (-6,-18);
\node [above] at (-12,-18) {$\mathtt{dec}$};

\node [above] at (7,-25) {$\mathsf{M}$};
\draw [rounded corners] (-3,-19) rectangle (17,1);
\draw [rounded corners, red, dashed,thick] (-4,-20) rectangle (18,2);

\draw [<-] (-18,10) -- (20,10);
\node [above] at (1,10) {$\mathtt{ebr}$};

\node [above] at (28,-25) {$\mathsf{S}$};
\node [scale=0.8] at (28,10)
{
\begin {tikzpicture} [scale=0.1]
\draw (0,-10) ellipse (7cm and 3cm);
\draw [white,fill=white] (0,-9.5) ellipse (7cm and 3cm);
\draw (0,10) ellipse (7cm and 3cm);
\draw (7,10) -- (7,-10);
\draw (-7,10) -- (-7,-10);
\draw [fill=yellow, even odd rule, yshift=-9cm, xshift=1cm] (0,0) rectangle (4,4) (1.6,2.4) -- (1.4,0.6) -- (2.6,0.6) -- (2.4,2.4) arc (-40:220:0.54);
\draw [fill=yellow, yshift=-9cm, xshift=1cm] (4,4) -- (4,5) arc (0:180:2) -- (0,4) -- (0.6,4) -- (0.6,5) arc (180:0:1.4) -- (3.4,4) -- (4,4);
\end {tikzpicture}
};

\draw [<-] (36,10) -- (45,10);
\node [above] at (40.5,10) {$\mathtt{ebr}$};

\node [above] at (56,-25) {$\mathsf{I}$};
\node [scale=0.4] at (56,2)
{
\begin {tikzpicture} [scale=0.5]
\draw [rounded corners] (0,0) rectangle (8,10);
\node at (1.5,6) {\includegraphics [width=1.1cm] {inputs/1101_1.jpg}};
\draw (0.5,4.5) rectangle (2.5,7.5);
\end {tikzpicture}
};
\node at (56,10) {$\textsc{br} \to \mathtt{ebr}$};
\node at (56,-5) {$\textsc{br} \to \mathtt{qr}$};

\draw [<-] (20,-5) -- (45,-5);
\node [below] at (32.5,-5) {$\mathtt{qr}$};

\end {tikzpicture}
\end {center}
\end {minipage}}
\end {center}
\caption {Architecture of the extension of the Match-On-Card technology to biometric identification.
The dotted red line indicates the location of the comparison.}
\label{fig:mocident}
\end {figure}

The verification step is processed as follows.
Suppose one wants to identify himself in the system.
A terminal captures the fresh biometrics, extracts a template, computes its quantization $\mathtt{qs}$ and sends them to the secure module.
Then, the module proceeds to a comparison between the fresh quantization and all enrolled quantizations $\mathtt{qr}$.
The \textsc{c} nearest quantizations, for some parameter \textsc{c} of the system, are the \textsc{c} potential candidates for the identification.
Then, the module queries the \textsc{c} corresponding (encrypted) templates to the database (by using the list of indices $\mathtt{ind}$ of those \textsc{c} nearest quantized versions $\mathtt{qr}$ of the enrolled templates). This gives the module the access to the set $\mathtt{sebr}$ of the \textsc{c} encrypted templates. The module decrypts them, and compares them with the fresh template $\mathtt{bs}$.
The module finally sends its response to the terminal: 1 if one of the enrolled templates is close enough to the fresh template, 0 otherwise.
Figure~\ref{fig:mocident} gives a graphical representation of the resulting architecture.

\textsc{n} denotes the size of the database (\textit{i.e.} the number of enrolled users), \textsc{q} the size of the quantizations, and \textsc{c} the number of indices asked by the card.
The ranges are $Range (\textsc{br}, \mathtt{ebr}, \mathtt{qr})$ $=$ \textsc{n}, $Range (\mathtt{rd}, \textsc{thr}, \mathtt{bs}, \mathtt{qs}, \mathtt{dec}) = 1$, and $Range (\mathtt{ind}, \mathtt{sebr}, \mathtt{sbr}) = \textsc{c}$.
The set $Fun$ of functions contains the extraction procedure $Extract$, the encryption and decryption procedures $Enc$ and $Dec$, the (non-invertible) quantization $Quant$ of the biometric templates, the comparison of the quantizations $QComp$, which takes as inputs two sets of quantizations and the parameter \textsc{c}, the selection of the encrypted templates $EGet$, and finally the matching $\mu$, which takes as arguments two biometric templates and the threshold~$\textsc{thr}$.

The biometric reference templates are enrolled by the issuer ($Has_\mathsf{I} (\textsc{br})$).
A verification process is initiated by the terminal \textsf{T} receiving as input a raw biometric data \texttt{rd} from the user \textsf{U}.
\textsf{T} extracts the fresh biometric template \texttt{bs} from \texttt{rd} using the function $Extract \in Fun$.
The architecture then contains, as other biometric systems, $ Receive_{\mathsf{T}, \mathsf{U}} (\{\}, \{\mathtt{rd}\}) $ and $ Compute_\mathsf{T} (\mathtt{bs} = Extract (\mathtt{rd}))$ and the $Dep_\mathsf{T}$ relation is such that $(\mathtt{bs}, \{\mathtt{rd}\}) \in Dep_\mathsf{T}$.
The user receives the final decision $\mathtt{dec}$ from the terminal: $Receive_{\mathsf{U}, \mathsf{T}} (\{\}, \{\mathtt{dec}\})$.
To sum up, the architecture is described as follows in the framework of Section~\ref{s:framework}:
\begin {align*}
A & {}^\mathsf{mi} := \big\{
Has_\mathsf{I} (\textsc{br}),
Has_\mathsf{U} (\mathtt{rd}),
Has_\mathsf{M} (\textsc{c}),
Has_\mathsf{M} (\textsc{thr}),
\\ &
Compute_\mathsf{I} (\mathtt{ebr} = Enc (\textsc{br})),
Compute_\mathsf{I} (\mathtt{qr} = Quant (\textsc{br})),
\\ &
Compute_\mathsf{T} (\mathtt{bs} = Extract (\mathtt{rd})),
Compute_\mathsf{T} (\mathtt{sebr} = EGet (\mathtt{ebr}, \mathtt{ind})),
\\ &
Compute_\mathsf{T} (\mathtt{qs} = Quant (\mathtt{bs})),
Compute_\mathsf{M} (\mathtt{ind} = QComp (\mathtt{qs}, \mathtt{qr}, \textsc{c})),
\\ &
Compute_\mathsf{M} (\mathtt{sbr} = Dec (\mathtt{sebr})),
Compute_\mathsf{M} (\mathtt{dec} = \mu (\mathtt{sbr}, \mathtt{bs}, \textsc{thr})),
\\ &
Receive_{\mathsf{S}, \mathsf{I}} (\{Attest_\mathsf{I} (\mathtt{ebr} = Enc (\textsc{br}))\}, \{\mathtt{ebr}\}),
Receive_{\mathsf{T}, \mathsf{U}} (\{\}, \{\mathtt{rd}\}),
\\ &
Receive_{\mathsf{T}, \mathsf{S}} (\{Attest_\mathsf{I} (\mathtt{ebr} = Enc (\textsc{br}))\}, \{\mathtt{ebr}\}),
Receive_{\mathsf{M}, \mathsf{T}} (\{\}, \{\mathtt{qs}\}),
\\ &
Receive_{\mathsf{M}, \mathsf{I}} (\{Attest_\mathsf{I} (\mathtt{qr} = Quant (\textsc{br}))\}, \{\mathtt{qr}\}),
Receive_{\mathsf{T}, \mathsf{M}} (\{\}, \{\mathtt{ind}\}),
\\ &
Receive_{\mathsf{M}, \mathsf{T}} (\{\}, \{\mathtt{sebr}, \mathtt{bs}\}),
Receive_{\mathsf{T}, \mathsf{M}} (\{\}, \{\mathtt{dec}\}),
\\ &
Trust_{\mathsf{T}, \mathsf{I}},
Trust_{\mathsf{M}, \mathsf{I}},
Trust_{\mathsf{T}, \mathsf{M}},
Verify_\mathsf{T}^{} (Attest_\mathsf{I} (\mathtt{ebr} = Enc (\textsc{br}))),
\\ &
Verify_\mathsf{T}^{} (\{Attest_\mathsf{M} (\mathtt{dec} = \mu (\mathtt{sbr}, \mathtt{bs}, \textsc{thr}))\}),
\\ &
Verify_\mathsf{M}^{} (Attest_\mathsf{I} (\mathtt{qr} = Quant (\textsc{br}))),
Verify_\mathsf{T}^{} (\{Attest_\mathsf{M} (\mathtt{sbr} = Dec (\mathtt{ebr}))\})
\big\}
\end {align*}
The issuer encrypts the templates and computes the quantizations, which is expressed by the dependencies:
$Dep_\mathsf{I}^\mathsf{mi}$ $:=$ \{($\mathtt{ebr}$, \{$\textsc{br}$\}),
($\mathtt{qr}$, \{$\textsc{br}$\})\}.
The terminal and module computations are reflected in the dependencies as well:
$Dep_\mathsf{T}^\mathsf{mi}$ $:=$ \{($\mathtt{bs}$, \{$\mathtt{rd}$\}),
($\mathtt{qs}$, \{$\mathtt{bs}$\})\},
($\mathtt{sebr}$, \{$\mathtt{bs}$, $\mathtt{ind}$\})\}.
The dependency relation of the module reflects its ability to decrypt the templates:
$Dep_\mathsf{M}^\mathsf{mi}$ := \{($\mathtt{ind}$, \{$\mathtt{qs}$, $\mathtt{qr}$, $\textsc{c}$\}),
($\mathtt{sbr}$, \{$\mathtt{sebr}$\}),
($\mathtt{dec}$, \{$\mathtt{sbr}$, $\mathtt{bs}$, $\textsc{thr}$\}),
($\textsc{br}$, \{$\mathtt{ebr}$\})\}.
The absence of such a relation in other dependencies prevents the corresponding components to get access to the plain references, even if they get access to the ciphertexts.

\subsection {Learning from the selected quantizations}

Let us now discuss the following point: the formalism of~Section~\ref{s:framework} is insufficient to consider the leakage of the sensitive biometric data stored inside the module.
In $A^\mathsf{mi}$, we would like that the terminal gets no access to the quantizations: $A^\mathsf{mi} \in Has^{none}_\mathsf{T} (\mathtt{qr})$.
It is indeed possible to derive $A^\mathsf{mi} \vdash Has^{none}_\mathsf{T} (\mathtt{qr})$, thanks to the (\textsf{HN}) rule.
According to the notations of~\cite{DBLP:conf/stm/AntignacM14}, where $Has_i (X)$ stands for $Has_i (X^{(1)})$ in this paper, we have:
\begin {center}
\AxiomC {$\nexists X: Dep_\mathsf{T} (\mathtt{qr}, X) \in A^{\mathsf{mi}}$}
\noLine
\UnaryInfC {$Has_\mathsf{T} (\mathtt{qr}) \not\in A^{\mathsf{mi}}$}
\AxiomC {$\nexists j, S: Receive_{\mathsf{T}, j} (S, \{\mathtt{qr}\}) \in A^{\mathsf{mi}}$}
\noLine
\UnaryInfC {$\nexists T: Compute_{\mathsf{T}} (\mathtt{qr} = T) \in A^{\mathsf{mi}}$}
\BinaryInfC {$A \nvdash Has_\mathsf{T} (\mathtt{qr})$}
\LeftLabel {\textsf{HN}}
\UnaryInfC {$A \vdash Has^{none}_\mathsf{T} (\mathtt{qr})$}
\DisplayProof
\end {center}
This corresponds to the intuition saying that quantizations are protected since they are stored in a secure hardware element.

However, an attack (described in \cite{DBLP:conf/iih-msp/BringerCS10}) shows that, in practice, quantizations can be learned if a sufficient number of queries to the module is allowed.
The attack roughly proceeds as follows (we drop the masks for sake of clarity).
The attacker maintains a $\textsc{n} \times \textsc{q}$ table (say $T$) of counters for each bit to be guessed.
All entries are initialized to 0.
Then it picks \textsc{q}-bits random vector $Q$ and sends it to the module.
The attacker observes the set of indices $\mathtt{ind} \subseteq [1, \textsc{n}]$ corresponding to the encrypted templates asked by the module.
It updates its table $T$ as follows, according to its query $Q$ and the response $\mathtt{ind}$:
for each $i \in [1, \textsc{n}]$ and $j \in [1, \textsc{q}]$, it decrements the entry $T[i][j]$ if $Q[j] = 0$, and increments it if $Q[j] =1$.
At the end of the attack, the \textsc{n} quantizations are guessed from the signs of the counters.

The number of queries made to the module is the crucial point in the attack above (and generally in other black-box attacks against biometric systems \cite{DBLP:conf/iih-msp/BringerCS10}).
Our extended model enables to introduce a bound on the number of actions allowed to be performed.
We now use this model to integrate such a bound in the formal architecture description.
Let $A^{\mathsf{mi\text-e}} (n)$ be the following architecture, for some $n \geq 1$:
\begin {align*}
A^{\mathsf{mi\text-e}} & (n) := \big\{
Has_\mathsf{I} (\textsc{br}),
Has_\mathsf{U}^{(n)} (\mathtt{rd}),
Has_\mathsf{M} (\textsc{c}),
Has_\mathsf{M} (\textsc{thr}),
\\ &
Compute_\mathsf{I}^{(n)} (\mathtt{ebr} = Enc (\textsc{br})),
Compute_\mathsf{I}^{(n)} (\mathtt{qr} = Quant (\textsc{br})),
\\ &
Compute_\mathsf{T}^{(n)} (\mathtt{bs} = Extract (\mathtt{rd})),
Compute_\mathsf{T}^{(n)} (\mathtt{sebr} = EGet (\mathtt{ebr}, \mathtt{ind})),
\\ &
Compute_\mathsf{T}^{(n)} (\mathtt{qs} = Quant (\mathtt{bs})),
Compute_\mathsf{M}^{(n)} (\mathtt{ind} = QComp (\mathtt{qs}, \mathtt{qr}, \textsc{c})),
\\ &
Compute_\mathsf{M}^{(n)} (\mathtt{sbr} = Dec (\mathtt{sebr})),
Compute_\mathsf{M}^{(n)} (\mathtt{dec} = \mu (\mathtt{sbr}, \mathtt{bs}, \textsc{thr})),
\\ &
Receive_{\mathsf{S}, \mathsf{I}}^{(n)} (\{Attest_\mathsf{I} (\mathtt{ebr} = Enc (\textsc{br}))\}, \{\mathtt{ebr}\}),
Receive_{\mathsf{T}, \mathsf{U}}^{(n)} (\{\}, \{\mathtt{rd}\}),
\\ &
Receive_{\mathsf{T}, \mathsf{S}}^{(n)} (\{Attest_\mathsf{I} (\mathtt{ebr} = Enc (\textsc{br}))\}, \{\mathtt{ebr}\}),
Receive_{\mathsf{M}, \mathsf{T}}^{(n)} (\{\}, \{\mathtt{qs}\}),
\\ &
Receive_{\mathsf{M}, \mathsf{I}}^{(n)} (\{Attest_\mathsf{I} (\mathtt{qr} = Quant (\textsc{br}))\}, \{\mathtt{qr}\}),
Receive_{\mathsf{T}, \mathsf{M}}^{(n)} (\{\}, \{\mathtt{ind}\}),
\\ &
Receive_{\mathsf{M}, \mathsf{T}}^{(n)} (\{\}, \{\mathtt{sebr}, \mathtt{bs}\}),
Receive_{\mathsf{T}, \mathsf{M}}^{(n)} (\{\}, \{\mathtt{dec}\}),
\\ &
Trust_{\mathsf{T}, \mathsf{I}},
Trust_{\mathsf{M}, \mathsf{I}},
Trust_{\mathsf{T}, \mathsf{M}},
Verify_\mathsf{T}^{(n)} (Attest_\mathsf{I} (\mathtt{ebr} = Enc (\textsc{br}))),
\\ &
Verify_\mathsf{T}^{(n)} (\{Attest_\mathsf{M} (\mathtt{dec} = \mu (\mathtt{sbr}, \mathtt{bs}, \textsc{thr}))\}),
\\ &
Verify_\mathsf{M}^{(n)} (Attest_\mathsf{I} (\mathtt{qr} = Quant (\textsc{br}))),
\\ &
Verify_\mathsf{T}^{(n)} (\{Attest_\mathsf{M} (\mathtt{sbr} = Dec (\mathtt{ebr}))\})
\big\}
\end {align*}
In addition to the dependence of $A^\mathsf{mi}$, the dependence relations indicates that the leakage is conditioned by a specific link mapping between the outsourced ciphertexts and the stored quantizations:
$Dep_\mathsf{T}^\mathsf{mi\text-e} (\mathtt{qr}, \{\mathtt{ind}^{(\textsc{n} \cdot \textsc{q})}, \mathtt{qs}^{(\textsc{n} \cdot \textsc{q})}\})$.
Furthermore, the module may learn the entire database \texttt{ebr} in a number of queries depending on the size of the database and the number of indices asked by the module:
$Dep_\mathsf{M}^\mathsf{mi\text-e} (\mathtt{ebr}, \{\mathtt{sebr}^{(\lceil \textsc{n} / \textsc{c} \rceil)}\})$.

\subsection {Strengthened variants of the architecture}

Now, based on some counter-measures of the attacks indicated in~\cite{DBLP:conf/iih-msp/BringerCS10}, we express several variants of the architecture $A^{\mathsf{mi\text-e}}$.
For each variant, the deductive rules $\mathcal{D}$ for the property language $\mathcal{L}_P$ are used to show that, for some conditions on the parameters, the quantizations \texttt{qr} are protected.

\subsubsection{Variant 1}
As a first counter-measure, the module could ask the entire database at each invocation.
It is rather inefficient, and, in some sense, runs against to initial motivation of its design.
However, this can be described within the language $\mathcal{L}_A$, and, in practice, can be manageable for small databases.
This architecture, denoted $A^\mathsf{mi\text-e1}$, is given by $A^\mathsf{mi\text-e} (n)$ for some $n \geq 1$, except that $Dep_\mathsf{T}^{\mathsf{mi\text-e1}}$ $:=$ $Dep_\mathsf{T}^{\mathsf{mi}}$.
It is now possible to prove that the quantizations are protected, even in presence of several executions of the protocols.
Since the relations $Dep_\textsf{T}$ no longer contains a dependence leading to \texttt{qr}, an application of (\textsf{HN}) becomes possible and gives the expected property.
\begin {center}
\AxiomC {$\nexists X: Dep_\mathsf{T} (\mathtt{qr}, X) \in A^{\mathsf{mi\text-e1}}$}
\noLine
\UnaryInfC {$Has_\mathsf{T}^{(n)} (\mathtt{qr}) \not\in A^{\mathsf{mi\text-e1}}$}
\AxiomC {$\nexists j: Receive_{\mathsf{T}, j}^{(n)} (S, \{\mathtt{qr}\}) \in A^{\mathsf{mi\text-e1}}$}
\noLine
\UnaryInfC {$\nexists T: Compute_{\mathsf{T}}^{(n)} (\mathtt{qr} = T) \in A^{\mathsf{mi\text-e1}}$}
\BinaryInfC {$\forall n: A \nvdash Has_\mathsf{T} (\mathtt{qr}^{(n)})$}
\LeftLabel {\textsf{HN}}
\UnaryInfC {$A \vdash Has^{none}_\mathsf{T} (\mathtt{qr})$}
\DisplayProof
\end {center}

\subsubsection {Variant 2}
In the precedent variant, the effect of the counter-measure is the withdrawal of the dependence relation.
We now consider architectures where such a dependency is still given, but where counter-measures are used to prevent a critical bound on the number of queries to be reached.

A first measure is to block the number of attempts the terminal can make.
The module can detect it and refuse to respond.
This architecture, denoted $A^{\mathsf{mi\text-e2}}$, is given by $A^{\mathsf{mi\text-e}} (\textsc{b})$, for some $\textsc{b} \ll \textsc{n} \cdot \textsc{q}$.
As a result, the $Has^{none}_i (\mathtt{qr})$ property can be derived.
In particular one must show that $A^{\mathsf{mi\text-e2}} \nvdash Has_\mathsf{T} (\mathtt{ind}^{(\textsc{n} \cdot \textsc{q})})$, in order to prevent the dependence rule \textsf{H5} to be applied.
\begin {center}
\AxiomC {$\nexists S: Receive_{\mathsf{T}, \mathsf{M}}^{(\textsc{b})} (S, \{\mathtt{ind}\}) \in A^{\mathsf{mi\text-e2}}$}
\AxiomC {$Has_\mathsf{T}^{(\textsc{b})} (\mathtt{ind}) \in A^{\mathsf{mi\text-e2}}$}
\AxiomC {$\textsc{b} < \textsc{n} \cdot \textsc{q}$}
\noLine
\BinaryInfC {$\nexists T: Compute_{\mathsf{T}}^{(\textsc{b})} (\mathtt{ind} = T) \in A^{\mathsf{mi\text-e2}}$}
\BinaryInfC {$A^{\mathsf{mi\text-e2}} \nvdash Has_\mathsf{T} (\mathtt{ind}^{(\textsc{n} \cdot \textsc{q})})$}
\DisplayProof
\end {center}
An application of \textsf{HN} enables to conclude.
\begin {center}
\AxiomC {$Dep_\mathsf{T}^\mathsf{mi\text-e2} (\mathtt{qr}, \{\mathtt{ind}^{(\textsc{n} \cdot \textsc{q})}\}) \in A^{\mathsf{mi\text-e2}}$}
\AxiomC {$Has_\mathsf{T}^{(\textsc{b})} (\mathtt{qr}) \not\in A^{\mathsf{mi\text-e2}}$}
\noLine
\UnaryInfC {$\nexists j: Receive_{\mathsf{T}, j}^{(\textsc{b})} (S, \{\mathtt{qr}\}) \in A^{\mathsf{mi\text-e2}}$}
\noLine
\BinaryInfC {$A^{\mathsf{mi\text-e2}} \nvdash Has_\mathsf{T} (\mathtt{ind}^{(\textsc{n} \cdot \textsc{q})})$ \qquad $\nexists T: Compute_{\mathsf{T}}^{(\textsc{b})} (\mathtt{qr} = T) \in A^{\mathsf{mi\text-e2}}$}
\UnaryInfC {$A^{\mathsf{mi\text-e2}} \nvdash Has_\mathsf{T} (\mathtt{qr}^{(1)})$}
\LeftLabel {\textsf{HN}}
\UnaryInfC {$A^{\mathsf{mi\text-e2}} \vdash Has^{none}_\mathsf{T} (\mathtt{qr})$}
\DisplayProof
\end {center}

\subsubsection{Variant 3}
In the precedent variant, the terminal cannot accumulate enough information since he cannot query the module enough times to derive a useful knowledge.
We now describe a variant where the terminal has no bound on the number of times it asks the module, but where the system is regularly reinitialised, so that the accumulated information becomes useless.

The leakage of the system runtime is dependent on some association between the quantizations \texttt{qr} and the encrypted database \texttt{ebr}; namely the association $\pi$ that maps the quantization $\mathtt{qr}[i] = Quant (\textsc{br}[{\pi (i)}])$ to the encrypted template from which it has been computed $\mathtt{ebr}[{\pi (i)}] = Enc (\textsc{br}[{\pi (i)}])$.
Once this mapping is changed, the information is cancelled.
For instance the database can be randomly permuted after \textsc{b} queries to the secure module.

Formally, this is caught by adding a $Reset$ primitive to the architecture.
Let $A^\mathsf{mi\text-e3}$ be the architecture defined as $A^\mathsf{mi\text-e3}$ $:=$ $A^\mathsf{mi\text-e2} \cup \{Reset\}$.
The semantics of the $Reset$ events ensures that no more than \textsc{b} values of \texttt{ind} will be gathered by the terminal for a fixed mapping.
The proof that $A^{\mathsf{mi\text-e3}} \vdash Has^{none}_\mathsf{T} (\mathtt{qr})$ is as the proof that $A^{\mathsf{mi\text-e2}} \vdash Has^{none}_\mathsf{T} (\mathtt{qr})$.

\section {Related works}
\label{s:rel}

Generally speaking, while the privacy of biometric data has attracted a lot of attention in the news (for instance, with the introduction of a fingerprint sensor in the new iphone) and among lawyers and policy makers\footnote{For example with a proposal adopted by the French Senate in May 2014 to introduce stronger requirements for the use of biometrics.}, it has not triggered such a strong interest in the computer science community so far.
Most studies in this area are done on a case by case basis and at a lower level than the architectures described here.
For instance, \cite{DBLP:conf/ispec/TangBCP08} proposes a security model for biometric-based authentication taking into account privacy properties -- including impersonation resilience, identity privacy or transaction anonymity -- and applies it to biometric authentication.
The underlying proofs rely on cryptographic techniques related to the ElGamal public key encryption scheme.
\cite{DBLP:journals/scn/KanakS14,DBLP:journals/tifs/LaiHP11,DBLP:journals/tifs/LaiHP11a} develop formal models from an information theoretic perspective relying on specific representations of biometric templates close to error correcting codes.

As far as formal approaches to privacy are concerned, two main categories can be identified: the qualitative approach and the quantitative approach.
Most proposals of the first category rely on a language which can be used to define systems and to express privacy properties.
For example process calculi such as the applied pi-calculus~\cite{DBLP:conf/popl/AbadiF01} have been applied to define privacy protocols~\cite{DBLP:conf/wote/DelauneKR10}.
Other studies~\cite{DBLP:conf/sp/BarthDMN06,becker:2010} involve dedicated privacy languages.
The main departure of the approach advocated in this paper with respect to this trend of work is that we reason at the level of architectures, providing ways to express properties without entering into the details of specific protocols.
Proposals of the second category rely on privacy metrics such as $k$-anonymity, $l$-diversity, or $\epsilon$-differential privacy~\cite{DBLP:conf/icalp/Dwork06} which can be seen as ways to measure the level of privacy provided by an algorithm.
Methods~\cite{DBLP:conf/sigmod/McSherry09} have been proposed to design algorithms achieving privacy metrics or to verify that a system achieves a given level of privacy.
These contributions on privacy metrics are complementary to the work described in this paper.
We follow a qualitative (or logical) approach here, proving that a given privacy property is met (or not) by an architecture.
As suggested in the next section, an avenue for further research would be to cope with quantitative reasoning as well, using inference systems to derive properties expressed in terms of privacy metrics.
 
Several authors \cite{gurses:2011,DBLP:conf/apf/Kerschbaum12,DBLP:conf/codaspy/Metayer13,mulligan:2012,DBLP:journals/tse/SpiekermannC09} have already pointed out the complexity of ``privacy engineering'' as well as the ``richness of the data space''\cite{gurses:2011} calling for the development of more general and systematic methodologies for privacy by design.
\cite{DBLP:conf/apf/Kerschbaum12,DBLP:conf/csfw/MaffeiPR13} point out the complexity of the implementation of privacy and the large number of options that designers have to face.
To address this issue and favour the adoption of these tools, \cite{DBLP:conf/apf/Kerschbaum12} proposes a number of guidelines for the design of compilers for secure computation and zero-knowledge proofs whereas \cite{DBLP:conf/uss/FournetKDL13} provides a language and a compiler to perform computations on private data by synthesising zero-knowledge protocols.
None of these proposals addresses the architectural level and makes it possible to get a global view of a system and to reason about its underlying trust assumption.
 
\section {Conclusion}
\label{s:conclusion}

This work is the result of a collaboration between academics, industry and lawyers to show the applicability of the privacy by design approach to biometric systems and the benefit of formal methods to this end.
Indeed, even if privacy by design becomes a legal obligation in the European Union~\cite{european-parliament:2014} its application to real systems is far from obvious.
We have presented in the same formal framework a variety of architectural options for privacy preserving biometric systems.
We also have introduced an extension of this formal framework in order to catch the leakage due to the system runtime.

One of the main advantages of the approach is to provide formal justifications for the architectural choices and a rigorous basis for their comparison.
Table~\ref{tab:comp} is a recap chart of the architectures reviewed in the first part of this paper.
One of the most interesting pieces of information is the trust assumptions which are highlighted by the model.
The first line shows that $A_\mathsf{ed}$ is the architecture in which the strongest trust in put in the terminal that does not have to trust any other component apart from the issuer and is able to get access to \texttt{br}.
Architecture $A_\mathsf{hsm}$ is a variant of $A_\mathsf{ed}$; it places less trust in the terminal that has to trust the hardware security module to perform the matching.
$A_\mathsf{hom}$ is the architecture in which the terminal is less trusted: it has to trust the issuer, the hardware security module and the server for all sensitive operations and its role is limited to the collection of the fresh biometric trait and the computation of the fresh template.
Architecture $A_\mathsf{moc}$ is similar to this respect but all sensitive operations are gathered into a single component, namely the smart card.
It should be clear that no solution is inherently better than the others and, depending on the context of deployment and the technology used, some trust assumptions may be more reasonable than others.
In any case, it is of prime importance to understand the consequences of a particular choice in terms of trust.

\begin {table}[ht]
\begin {center}
\begin {tabular}{|c|c|c|c|c|}
\hline
Arch. & Computations & \multicolumn{2}{c|}{Template protection} & Trust relations \\
 & & \multicolumn{2}{c|}{} & \\
 & & Components & Components & \\
 & Location of & accessing the & accessing & \\
 & the matching & references \texttt{br} & the query \texttt{bs} & \\
\hline
$A_\mathsf{ed}$ & \textsf{T} & \textsf{I}, \textsf{T} & \textsf{T} & (\textsf{T}, \textsf{I}) \\
$A_\mathsf{hsm}$ & \textsf{M} & \textsf{I}, \textsf{M} & \textsf{T}, \textsf{M} & (\textsf{T}, \textsf{I}), (\textsf{T}, \textsf{M}) \\
$A_\mathsf{hom}$ & \textsf{S} & \textsf{I} & \textsf{T} & (\textsf{T}, \textsf{I}), (\textsf{T}, \textsf{M}), (\textsf{T}, \textsf{S}) \\
$A_\mathsf{moc}$ & \textsf{M} & \textsf{M} & \textsf{T}, \textsf{M} & (\textsf{T}, \textsf{M}) \\
\hline
\end {tabular}
\end {center} \scriptsize
Components are: user \textsf{U}, terminal \textsf{T}, server \textsf{S}, secure module \textsf{M} (used as a generic name for a hardware security module or a card \textsf{C}), issuer~\textsf{I}.

A trust relation $(i, j)$ means that component $i$ trusts component $j$.

\medskip

\caption{Comparison between architectures}
\label{tab:comp}
\end {table}

A benefit of the formal approach followed in this paper is that it can provide the foundations for a systematic approach to privacy by design.
A proof of concept implementation of a system to support designers in their task has been proposed in~\cite{DBLP:conf/ifiptm/AntignacM15}.
In this system, the user can introduce his privacy and integrity requirements (as well as any requirements imposed by the environment such as the location of a given operation on a designated component) and choose different options for the distribution of the operations and the trust assumptions.
When an architecture has been built, the system can try to verify the required properties with or without the help of the designer.

As stated above, we focused on the architectural level.
As a result, we do not cover the full development cycle.
Preliminary work has been done to address the mapping from the architecture level to the protocol level to ensure that a given implementation, expressed as an applied pi-calculus protocol, is consistent with an architecture \cite{DBLP:conf/fps/TaA14}.
As far as the formal approach is concerned, it would also be interesting to study how it could be used in the context of future privacy certification schemes.
This would be especially interesting in the context of the European General Data Protection Regulation~\cite{european-parliament:2014} which promotes not only privacy by design but also privacy seals.



\bibliographystyle{plain}
\bibliography{biopriv}

\end{document}